\documentclass[lettersize,journal]{IEEEtran}
\usepackage{amsmath,amssymb,dsfont,stfloats,color,url,bbm}
\usepackage[pdftex]{graphicx}
\usepackage{subfigure}
\usepackage{cite}
\usepackage{xcolor}
\usepackage{setspace}

\usepackage{resizegather}

\DeclareGraphicsExtensions{.eps,.pdf,.png,.jpg,.gif,.jpeg,.pstex}

\newtheorem{rem}{Remark}

\newfont{\bbb}{msbm10 scaled 700}

\newfont{\bb}{msbm10 scaled 1100}
\newcommand{\CC}{\mbox{\bb C}}
\newcommand{\PP}{\mbox{\bb P}}

\newcommand{\EE}{\mbox{\bb E}}

\newcommand{\HH}{\mbox{\bb H}}
\newcommand{\SSS}{\mbox{\bb S}}
\newcommand{\UU}{\mbox{\bb U}}

\newcommand{\BB}{\mbox{\bb B}}
\renewcommand{\AA}{\mbox{\bb A}}

\newcommand{\yy}{\mathbbm{y}}
\newcommand{\xx}{\mathbbm{x}}
\newcommand{\zz}{\mathbbm{z}}
\newcommand{\sss}{\mathbbm{s}}

\newcommand{\hh}{\mathbbm{h}}
\newcommand{\uu}{\mathbbm{u}}
\newcommand{\vvv}{\mathbbm{v}}

\newcommand{\av}{{\bf a}}

\newcommand{\hv}{{\bf h}}

\newcommand{\qv}{{\bf q}}
\newcommand{\rv}{{\bf r}}
\newcommand{\sv}{{\bf s}}

\newcommand{\uv}{{\bf u}}
\newcommand{\wv}{{\bf w}}
\newcommand{\vv}{{\bf v}}

\newcommand{\yv}{{\bf y}}
\newcommand{\zv}{{\bf z}}
\newcommand{\zerov}{{\bf 0}}
\newcommand{\onev}{{\bf 1}}

\newcommand{\Dm}{{\bf D}}

\newcommand{\Fm}{{\bf F}}
\newcommand{\Gm}{{\bf G}}
\newcommand{\Hm}{{\bf H}}
\newcommand{\Id}{{\bf I}}

\newcommand{\Ym}{{\bf Y}}
\newcommand{\Zm}{{\bf Z}}

\newcommand{\Cc}{{\cal C}}

\newcommand{\Ec}{{\cal E}}

\newcommand{\Gc}{{\cal G}}

\newcommand{\Nc}{{\cal N}}

\newcommand{\Sc}{{\cal S}}

\newcommand{\Uc}{{\cal U}}

\newcommand{\nuv}{\hbox{\boldmath$\nu$}}
\newcommand{\muv}{\hbox{\boldmath$\mu$}}
\newcommand{\zetav}{\hbox{\boldmath$\zeta$}}
\newcommand{\phiv}{\hbox{\boldmath$\phi$}}

\newcommand{\xiv}{\hbox{\boldmath$\xi$}}

\newcommand{\Gammam}{\hbox{\boldmath$\Gamma$}}

\newcommand{\Sigmam}{\hbox{\boldmath$\Sigma$}}

\newcommand{\Thetam}{\hbox{\boldmath$\Theta$}}

\newcommand{\Xim}{\hbox{\boldmath$\Xi$}}

\newcommand{\diag}{{\hbox{diag}}}

\newcommand{\trace}{{\hbox{tr}}}

\newcommand{\eqdef}{\stackrel{\Delta}{=}}

\newcommand{\herm}{{\sf H}}

\newcommand{\transp}{{\sf T}}

\newcommand{\SINR}{{\sf SINR}}
\newcommand{\SNR}{{\sf SNR}}

\usepackage{stmaryrd} 

\usepackage{hyperref}
\hypersetup{
    bookmarks=true,         
    unicode=false,          
    pdftoolbar=true,        
    pdfmenubar=true,        
    pdffitwindow=false,     
    pdfstartview={FitH},    
    pdfnewwindow=true,      
    colorlinks=true,       
    linkcolor=red,          
    citecolor=green,        
    filecolor=blue,      
    urlcolor=blue           
}

\begin{document}
\title{User-Centric Cell-Free Wireless Networks for 6G: Communication Theoretic Models and Research Challenges}

\author{Fabian G\"ottsch$^*$, Giuseppe Caire$^*$, Wen Xu$^\dagger$, Martin Schubert$^\dagger$ \thanks{$^*$ Faculty of EECS, Technische Universit\"at Berlin, Einsteinufer 25, 10587 Berlin, Germany} \thanks{$^\dagger$ Huawei Technologies D\"usseldorf GmbH, Riesstra{\ss}e 25, 80992 Munich, Germany}}

\date{\today}
\maketitle

\begin{abstract}
This paper presents a comprehensive communication theoretic model for the physical layer of a cell-free user-centric network, formed by 
user equipments (UEs), radio units (RUs), and decentralized units (DUs), uniformly spatially distributed over a given coverage area. 
We consider RUs equipped with multiple antennas, and focus on the regime where the 
UE, RU, and DU densities are constant and therefore the number of such nodes grows with the coverage area. 
A system is said scalable if the computing load and information rate at any node in the network converges to a constant as the network size (coverage area) grows to infinity. This imposes that each UE must be processed by a (user-centric) finite-size cluster of RUs, and that such cluster processors are dynamically allocated to 
the DUs (e.g., as software defined virtual network functions) in order to achieve a balanced computation load. 
We also assume that the RUs are connected to the DUs through a packet switching network, in order to achieve adaptive routing and load balance. 
For this model, we define in details the dynamic cluster formation and uplink pilot allocation. As a consequence of the pilot allocation and the scalability constraint, 
each cluster processor has a partial view of the network channel state information. We define the condition of ``ideal partial CSI'' when the channel vectors that can be estimated are  perfectly known (while the ones that cannot be estimated are not know at all). We develop two attractive cluster-based linear receiver schemes for the uplink, and an uplink-downlink duality that allows to reuse such vectors as precoders for the downlink. Finally, we show that exploiting the channel antenna correlation structure arising from a geometrically consistent model for directional propagation (which is well-motivated for short distance semi-line of sight propagation typical of dense wireless networks), and performing a channel subspace projection of the uplink pilot field, the pilot contamination effect arising from pilot reuse across the network can be effectively reduced to provide only marginal degradation with respect to the ideal partial CSI case. 
Several system aspects such as initial acquisition of the UEs, UE-RU association, and distributed scheduling for fairness and load balance between uplink down 
downlink are briefly discussed and identified as challenging research topics for further investigation. 
\end{abstract}


\section{Introduction}  \label{sec:intro}

The ever-growing demand for wireless data and ubiquitous broadband connectivity is pushing industry and standardization bodies to 
develop and release new generations of wireless systems designed to meet such demands. 
Since the very beginning, cellular has been the dominant architecture paradigm for 
outdoor wireless networks (e.g., see \cite{goldsmith2005wireless} and references therein). Namely, 
a given coverage area $A$ is partitioned into cells, and user equipments (UEs) in any given cell are uniquely associated to the corresponding base station (BS),\footnote{For simplicity of exposition we do not distinguish here between cells and sectors, which are conceptually equivalent when each sector handles its own UEs individually.} which 
implements the full stack of the radio access protocol. 
BSs are connected via a backhaul network to the rest of the system, and eventually through some gateway to 
the Internet. The area capacity in bit/s/m$^2$ is mainly driven by the following three factors: 1) cell density; 2) system bandwidth; 3) physical layer (PHY) and multiaccess schemes.

Taking the Gaussian capacity formula as a rule of thumb to represent the efficiency of the underlying PHY, 
the area capacity of a cellular system can be expressed as
\[ C = \lambda_a \times W \times \eta \log(1 +  \SINR) + {\rm const.} \]
where $\lambda_a$ is the cell density (number of BS per unit area), $W$ is the system bandwidth in Hz, 
$\eta \log(1 +  \SINR)$ is the sum spectral efficiency per cell supported by the PHY, and where
the constant term accounts for the non-ideal implementation and technology effects.
The cell spectral efficiency  can be further broken down into 
the pre-log factor $\eta$,  usually referred to as the {\em multiplexing gain}, and the  term $\log(1 + \SINR)$, corresponding to the capacity in bit 
per channel use of a Gaussian channel with given Signal to Interference plus Noise Ratio (SINR). 
This expression corresponds to a PHY scheme able to support $\eta$ virtual parallel Gaussian channels in the spatial domain, i.e., 
using multiple  antenna MIMO technology (e.g., see \cite{marzetta2016fundamentals} and references therein).
In addition, the term $\SINR$ expresses the average {\em Signal to Interference plus Noise Ratio} of a typical user randomly located in the cell area, 
where averaging is with respect to large-scale effects (distance dependent pathloss and shadowing/blocking effects), as well as small-scale effects (the fading due to multipath propagation). Taking the average SINR inside the log function yields an upper bound due to Jensen's inequality and the concavity of the logarithm. 
A refined analysis of the area capacity can be obtained using stochastic geometry approaches (e.g., see \cite{baccelli2009stochastic,haenggi2012stochastic,chen2016area}). In addition, it is often useful to characterize the system performance in terms of the {\em Cumulative Distribution Function} (CDF) of the per-user rate, instead of the (average) area capacity. 
In this case, one should consider the rate of a given user randomly placed in the cell, with averaging with respect to the small-scale fading and conditioning
on the large-scale effects. As a conditional average, this rate is a random variable, and the corresponding CDF yields the per-user rate 
distribution (e.g., see  \cite{bursalioglu2018fog,bayat2018achieving}). Nevertheless, for the sake of the discussion in this introduction section, the above simple formula is sufficient to capture the main  factors influencing the system performance.  

Historically, the spectrum allocated to cellular systems has steadily grown with the successive system generations. However, the availability of  ``beachfront spectrum'' 
-- namely, licensed spectrum below 6 GHz with broad geographic support -- is very limited \cite{andrews2016we}. 
Various windows of a few hundreds of MHz are available in different geographic regions, and systems must accommodate for bandwidth aggregation and dynamic spectrum allocation in order to scavenge such scarce available spectrum. 
Meanwhile, the available spectrum in higher frequency bands (e.g., 7-11 GHz and 24-54 GHz ) is certainly more plentiful, 
although still quite unproven for cellular outdoor and mobile 
communications as such frequencies struggle to penetrate walls and other blocking objects. 

Given that $W$ cannot grow significantly (at least in the spectrum below 6 GHz), the area capacity can be increased by letting the product 
$\lambda_a \times \eta$ grow, provided that the SINR  of the virtual per-user channels does not collapse to too low values. 
The increase of the multiplexing gain $\eta$ is achieved through multiuser MIMO (MU-MIMO) technology. 
In a conventional cellular system, each BS is equipped with an array of $M$ antenna elements, providing effectively a vector channel (namely, a vector Gaussian multiple-access channel (MAC) in the uplink (UL), and a vector Gaussian broadcast channel (BC) in the downlink (DL)). 
As long as the rank of the channel matrix between the BS antenna array and the served users has rank not smaller than some integer $d$ 
and is known at the BS side wish sufficient accuracy, MU-MIMO receivers (for the UL) and precoders (for the DL) can be designed in order to 
support $d$  virtually non-interfering data streams in both the UL and the DL.  
In general, the multiplexing gain $\eta$ is less than $d$ since 
other effects must be taken into account, and in particular, the overhead incurred by estimating the UL and DL channel matrix.

Since the first information theoretical studies \cite{caire2003achievable,Viswanath-Tse-TIT03,Weingarten-Steinberg-Shamai-TIT06,Caire-Jindal-Kobayashi-Ravindran-TIT10} to the inclusion in recent wireless standards \cite{3gpp38211,khorov2018tutorial,qu2019survey}, 
MU-MIMO is arguably one of the key transformative ideas that have shaped the last 15 years of theoretical research and practical technology development. 
 A successful related concept is {\em massive MIMO} \cite{marzetta2016fundamentals,marzetta2010noncooperative}. 
This is based on the key idea that, thanks to channel reciprocity and time-division duplexing (TDD) operations,  
an arbitrarily large number $M$ of BS antennas can be trained by a finite number of UEs 
using a finite-dimensional UL pilot field of $\tau_p \geq d$ UL symbols per coherence block.  

In TDD systems, UL and DL operate on the same frequency band. The channel reciprocity condition
is verified if the UL and DL slots occur within an interval significantly shorter than a channel coherence time, and if the Tx/Rx hardware of the BS radio are
calibrated. Hardware calibration has been widely demonstrated in practical testbeds, and can be achieved either using particular RF design solutions 
(e.g., see \cite{benzin2017internal}) or using over-the-air calibration (e.g., see \cite{rogalin2014scalable}). 
As far as the channel coherence time is concerned, a typical mobile user 
at carrier frequency between 2.0 and 3.7 GHz incurs Doppler typical spreads of $\sim 100$ Hz corresponding to channel coherence times of $\sim 10$ ms.  
For example, with TDD slots of 1 ms (corresponding to a subframe of a 10ms frame of 5GNR), 
we are well in the range for which the channel can be considered time-invariant 
over a UL/DL cycle. For faster moving users or higher carrier frequencies, the phenomenon of ``channel aging'' \cite{truong2013effects} between the UL and the DL slot 
cannot be neglected any longer. Specific channel estimation and short-term prediction (across the UL/DL slot) have been investigated 
in the literature, in particular for mmWave bands where the propagation happens to be mainly along the line-of-sight (LOS) and a few discrete reflection paths. 
However, a thorough discussion of these aspects goes beyond the scope of the present paper.

In this work we assume that  the channel is exactly constant over a time-frequency tile of $T$ channel uses in the 
time-frequency plane, where $T = T_c \times W_c$ and $T_c$ denotes the channel coherence time (in s) and $W_c$ denotes the channel coherence bandwidth (in Hz). 
For example, a coherence block may coincide (roughly) with a so-called Resource Block (RB) of 5GNR, consisting of 12 subcarriers $\times$ 14 OFDM symbols in time, for a total of $T = 168$ channel uses.  

Under such simplifying assumption, the resulting multiplexing gain is 
equal to $\eta = d (1 - \tau_p/T)$ for $d \leq \min\{\tau_p, M\}$.   
With massive MIMO (i.e., for $M$ sufficiently large) and per-cell processing  \cite{marzetta2010noncooperative}, eventually the number of UL and DL 
data streams $d$ is  limited by the pilot dimension $\tau_p$ and by the coherence block length $T$. 
In particular, provided that $M > T/2$, by letting $d = \tau_p$ and maximizing $\eta$ with respect to $\tau_p$, we find that the maximum 
multiplexing gain is achieved by letting $\tau_p = T/2$ and yields $\eta = T/4$. This depends only on 
the propagation and mobility characteristics of the physical channel, which determine $T_c$ and $W_c$ and therefore  $T$.

From the above discussion, it follows that the only other way to increase cellular capacity is cell densification, i.e., 
increasing $\lambda_a$. However, also $\lambda_a$ cannot increase indefinitely \cite{andrews2016we}, at least as far as a classical cellular 
architecture is adopted.  When the cell density increases (and consequently the cell size reduces), several problems emerge: 
\begin{itemize}
\item The frequency of handovers increases, with the consequent increase of protocol overhead and delay jitter.
\item The inter-cell interference increases to unbearable levels, yielding a too low SINR. 
This is due to the fact that even if each BS reduces its transmit power as a consequence of 
the smaller cell size, when the cells are very small, the propagation tends to become closer and closer to free-space conditions, for which the pathloss exponent is
small. In large cells with tower-mounted BSs the cell boundaries are enforced by  vertical tilting of the BS antenna radiation pattern in the elevation direction.  
In contrast, small-cell BSs are typically not mounted high on the ground and controlling inter-cell interference by tilting the radiation pattern in the elevation direction is more difficult.  
\item Specifically in MU-MIMO systems, the total number of users in the system is much larger than the 
UL pilot dimension $\tau_p$. Hence, UL pilots are reused across the network. This  pilot reuse yields the so-called {\em pilot contamination} effect, i.e., 
a coherently beamformed inter-cell interference which does not vanish even  if the number of antennas per BS $M$ grows to infinity \cite{huh2012achieving,hoydis2013massive}. Such effect is more and more dominant when the cell size shrinks, while the number of antennas 
per BS $M$ remains large. 
\end{itemize}
In view of the above problems,  the simple increase of the cell density $\lambda_a$, 
while insisting on a conventional cellular architecture with unique UE-to-BS association and per-cell processing, 
leads to a number of problems such that, eventually, the user rate will degrade and eventually the cellular area capacity 
reaches a plateau over which any further increase in the cell density yields only an increase in the cost of the infrastructure, 
without any benefit in area capacity. 

\subsection{Beyond the cellular architecture}

After realizing the intrinsic limitations of the cellular architecture with per-BS non-cooperative processing,  a flurry of works advocating the 
{\em joint processing} of spatially distributed infrastructure antennas has appeared.  
This idea can be traced back to the work of Wyner \cite{wyner1994shannon}, and in the communication theoretic and information theoretic literature 
has  been ``re-marketed''  several times under different names with slightly different nuances, 
such as {\em coordinate multipoint} (CoMP) \cite{jungnickel2014role,huh2011network},  
{\em cloud (or centralized) radio access network} (CRAN) \cite{checko2014cloud,park2013joint,aguerri2019capacity}, and, more recently, 
as  {\em cell-free massive MIMO} \cite{ngo2017cell,zhang2019cell,ngo2015cell}.
An excellent recent review of this vast literature is given in \cite{9336188}.\footnote{For the sake of precision, it should be said that CRAN is a network architecture, while CoMP is an LTE feature. However, 
the differences between these terms in the way they have been used in the theoretical literature is blurred. For example, 
there is no major conceptual difference between CRAN and CoMP with ``full joint processing'' from a theoretic viewpoint. 
In both cases, the signal to (resp., from) multiple spatially distributed users sent from (resp., received by) multiple distributed infrastructure antennas 
is jointly precoded (resp., jointly decoded) at a central processor.  Also, there is no conceptual difference between a cell-free system with fully centralized processing of all antenna sites and users a single giant cell with distributed antennas implemented via the CRAN architecture. 
Since this paper illustrates the problem from a theoretical viewpoint, 
we shall not discuss further the classification of alternative system proposals from a practical implementation or standard specification viewpoint.}

Advantages of this approach are the mitigation of pathloss and blocking, 
introducing proximity between the remote radio units (RUs) and UEs and macro-diversity, and the (obvious) elimination 
of inter-cell interference, by providing a single giant RU cluster. 
Both points, though, must be carefully discussed. First, deploying a number of RUs much larger than the number of UEs
is practically problematic, very costly, and often infeasible especially for outdoor systems.  
Then, the joint antenna processing across the whole network
does not eliminate the problem of a limited UL pilot dimension $\tau_p \ll K$ and therefore the consequent pilot contamination. 
Finally, global processing and optimization/allocation of pilots and transmit power across the network 
yield a non-scalable architecture. 

Current research has somehow agreed on a middle-point between conventional per-cell processing and centralized processing of all the RUs 
in the network. In particular, we may imagine a network formed by $K$ UEs, $L$ RUs and $D$ {\em Decentralized Units} (DUs), where typically $K > L > D$. 
The number of antennas per RU is denoted by $M$, and in the massive MIMO regime we have $ML > K$ (the total number of antennas is larger than the 
number of simultaneously active users). The RUs are connected to the DUs via a routing fronthaul network. Every user $k$ in the system is associated to a cluster
$\Cc_k \subseteq [L]$ of surrounding RUs,\footnote{The set of the first positive $N$ integers is given by $[N] = \{1, 2, \dots, N\}$.} and each RU $\ell$ serves a subset of users $\Uc_\ell \subseteq [K]$, given by the users $k$ such that $\ell \in \Cc_k$. Clusters are ``user-centric'', i.e., each user is associated to its own cluster, and different users may have different (partially overlapping) clusters. 
Each cluster $\Cc_k$ is jointly processed at some DU, which has to collect all signals from/to all the RUs $\ell \in \Cc_k$. 
The user-centric cluster processors are dynamically allocated to the DUs and are typically run in software over general purpose hardware. 
As users move through the network, their clusters evolve and ``follow'' them. Correspondingly, the cluster processors (virtualized network functions) 
are dynamically allocated to different DUs, in order to achieve load balancing in the routing fronthaul network. 

For such a system, we adopt the definition of scalability given by Bj\"ornson and Sanguinetti \cite{9064545}, informally recalled as follows: 
consider a network as described above, with covering area $A$ on the plane and UE, RU, and DU densities $\lambda_{\rm ue}, \lambda_{\rm ru}, \lambda_{\rm du}$, respectively, such that we have $K = \lambda_{\rm ue} A$, $L = \lambda_{\rm ru} A$ and $D = \lambda_{\rm du} A$. 
An architecture is said to be {\em scalable} if the complexity of the involved signal processing 
functions and the data rate conveyed at each point in the network converge to some constant values as $A \rightarrow \infty$.

\subsection{Related work}

The first works on cell-free massive MIMO consider a system, in which each UE is connected to all RUs \cite{ngo2017cell, 7917284, 8845768}, investigating different uplink (UL) and downlink (DL) methods. The UL with four levels of cooperation among the RUs is studied in \cite{8845768}, and it is shown that the cell-free architecture can outperform conventional cellular massive MIMO and small cells in terms of per user spectral efficiency (SE). In addition, it is shown that minimum
mean-square error (MMSE) combining is needed, as maximum ratio combining (MRC) is not able to outperform the compared network types. The presented methods however are not scalable, since all RUs are connected to all UEs. 
This issue is addressed in \cite{9064545, 8761828} by introducing the formation of finite size clusters such that each UE is only connected to a subset of all RUs. A distributed UL implementation with global large-scale fading decoding (LSFD) is presented in \cite{demir2021cellfree}, where after forming UE and RU clusters, the global symbol estimate of an RU cluster for a specific UE is formed by a weighted sum of the local estimates with the so-called {\em LSFD weights}. Distributed DL power allocation at the RUs and a combination of cell-centric and user-centric clusters are investigated in \cite{8761828}.
More user-centric approaches are proposed in \cite{8000355, 8097026, bursalioglu2018fog}. The effect of the number of UEs served by an RU on the UL and DL rates is investigated in \cite{8000355} for different channel estimation techniques. 
The energy efficiency of two distinct RU selection schemes is analyzed and compared to collocated massive MIMO in \cite{8097026}, considering backhaul power consumption, the number of RUs, and the number of antennas per RU.
The spectral efficiency and outage probability using stochastic geometry are studied in \cite{bursalioglu2018fog} in a {\em fog massive MIMO} system, where UEs with coded UL pilots seamlessly migrate from one RU to another with very low RU association latency.

Focusing on the DL, zero-forcing (ZF) beamforming and maximum ratio transmission (MRT) are studied in cell-free systems and compared to a small-cell architecture in \cite{7917284}, where ZF in the cell-free system outperforms the other methods for most UEs. Different local precoding methods extending the ``simple'' ZF are proposed in \cite{9069486} and achieve larger rates compared to the simple ZF in the deployed system with independent Rayleigh fading channels. 
Exact UL-DL duality for the ``true'' {\em use-and-then-forget (UatF)} bounds is studied in \cite{9064545}, assuming the knowledge of all required quantities in the SINR expressions also for not associated RU-UE pairs. Additionally, three scalable precoding schemes (partial MMSE, local partial MMSE and MRT) are investigated for two network topologies with the same number of total RU antennas in the system but with a different level of antenna concentration. 
As mentioned before, a comprehensive overview of this literature is given in the tutorial monograph \cite{9336188}.

In \cite{goettsch2021impact}, a model called {\em ideal partial CSI} is proposed, where each RU only has channel information of the associated UEs. 
This is due to the fact that, in the cluster association process, only the RUs belonging to cluster $\Cc_k$ are aware of the association between user $k$ and its allocated UL pilot signal.  The proposed combining method in \cite{goettsch2021impact} aims at maximizing the so-called ``optimistic ergodic rate'', i.e., the achievable rate when the UE receiver knows the useful signal term and the interference plus noise power. 
This work has been extended in \cite{WCNC2021} by considering also the DL and by showing that almost symmetric UL and DL rates can be achieved 
with a duality concept based on partial channel knowledge and ``nominal'' SINRs, i.e., assumed SINRs based on available channel information. 
These models and results will be reviewed in details in the following sections. 

\subsection{Contributions}

In this overview work we present in details the model of a user-centric cell-free scalable system based on TDD reciprocity and MU-MIMO (distributed) precoding
given in  \cite{goettsch2021impact,WCNC2021}. We consider  UL combining and DL precoding schemes, based on ideal partial CSI. 
We also provide an SINR UL/DL duality based on a definition of SINR that depends only on the partial CSI (i.e., what can be effectively measured at each cluster processor: channel estimates are available for associated UE-RU pairs instead of ideal channel knowledge), and demonstrate that such duality yields almost identical per user SE in terms of the actual SINRs and corresponding ergodic rates. 
Although not being the focus of this paper, different channel subspace and covariance estimation techniques are investigated in various works (see e.g. \cite{8067658, 1315936, 6126034}).  In this work, we assume that some user channel covariance estimation technique is used such that 
the dominant channel subspace can be reliably estimated. Based on this knowledge, we consider a simple approach to pilot decontamination based on {\em dominant subspace projection}. It will be demonstrated by simulation, that such approach is sufficient to closely approximate the
performance under the ideal partial CSI assumption, i.e., after subspace projection the estimated channels yield a per-user SE which is essentially equal (up to a small degradation) to the one obtained with ideal (but partial) channel knowledge. 
We also discuss the differences and similarities of the proposed UL combining and DL precoding schemes with respect to the current state of the art. 

Finally, we provide a number of interesting points for further research both at the PHY level (signal processing for channel estimation), 
and at the MAC/resource allocation level (signaling and algorithms for dynamic cluster formation, resource allocation and fairness scheduling).

\section{System model}  \label{sec:system}

\begin{figure}[h!]
	\centerline{\includegraphics[trim={140 130 120 82}, clip, width=.4\linewidth]{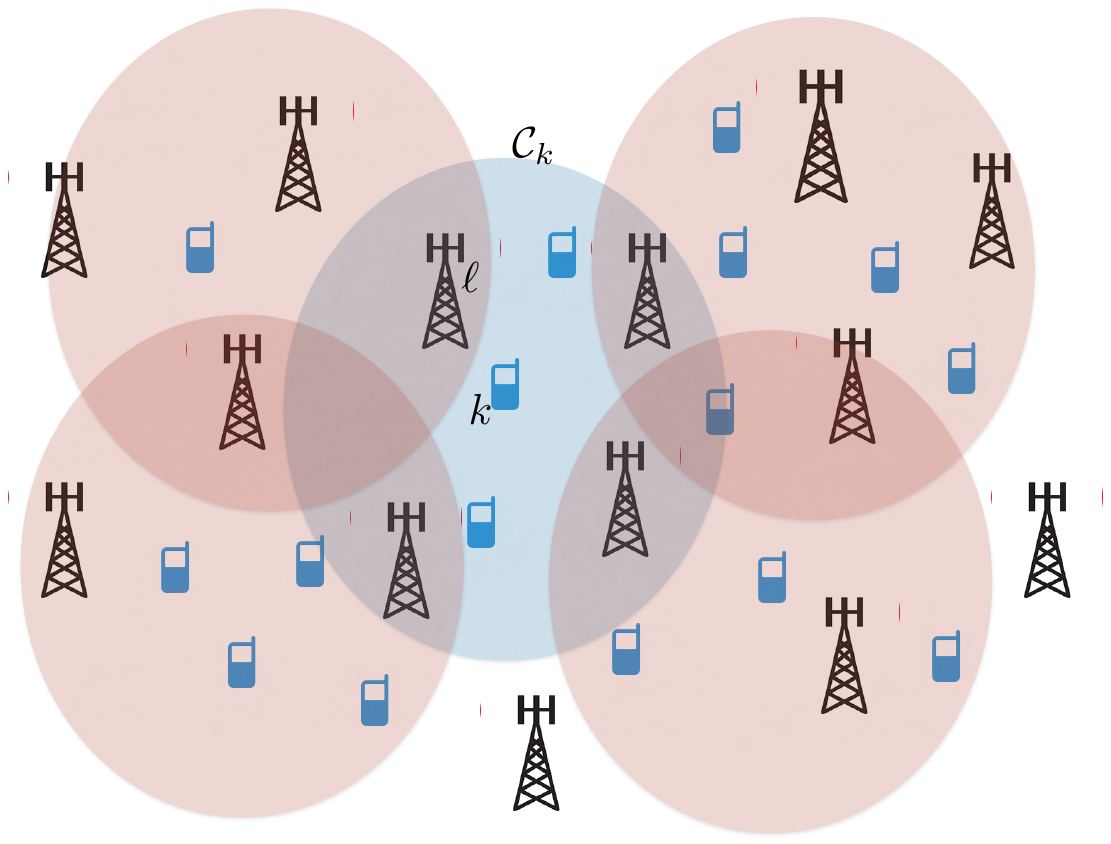} \includegraphics[trim={100 160 140 170}, clip, width=.54\linewidth]{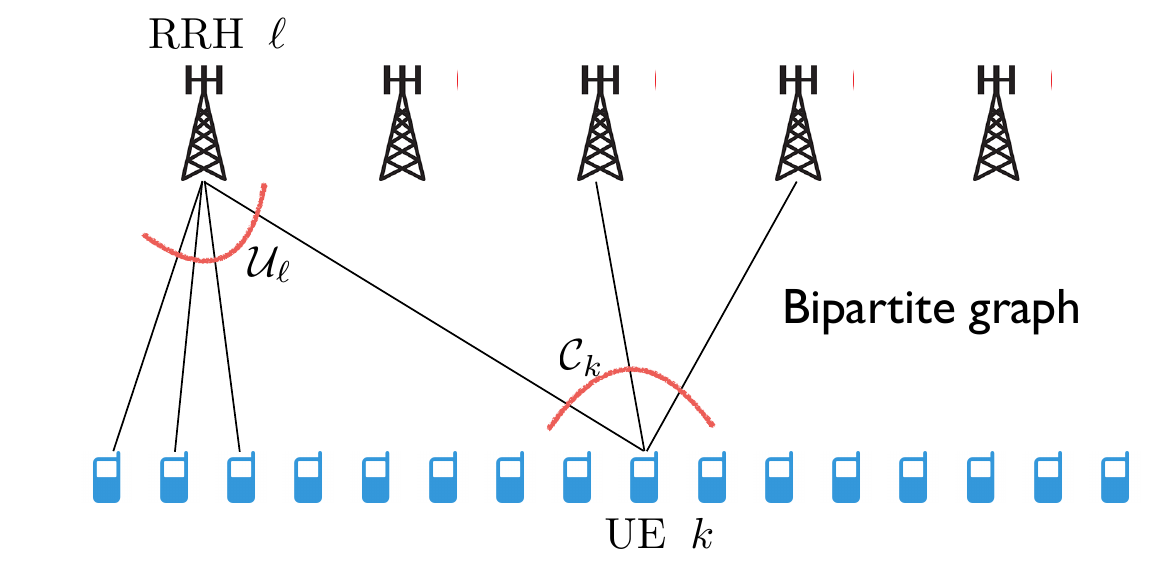}}
	\vspace{-.3cm}
	\caption{An example of dynamic clusters and the UE-RU association graph. The graph contains a UE-RU edge $(k,\ell)$ for all 
	$k \in [K]$ and $\ell \in [L]$ such that  $k \in \Uc_\ell$ and $\ell \in \Cc_k$.}
	\label{clusters}
\end{figure}

We consider a cell-free wireless network with $L$ RUs, each equipped with $M$ antennas, 
and $K$ single-antenna UEs. Both RUs and UEs are distributed on a squared region on the 2-dimensional plane. 
As a result of the cluster formation process (to be specified later), each UE $k$ is associated with a cluster $\Cc_k \subseteq [L]$ of RUs 
and each RU $\ell$ has a set of associated UEs $\Uc_\ell \subseteq [K]$. The UE-RU association is described by 
a bipartite graph $\Gc$ with two classes of nodes (UEs and RUs) such that the neighborhood of UE-node $k$ is $\Cc_k$ 
and the neighborhood of RU-node $\ell$ is $\Uc_\ell$. An example is given in Fig.~\ref{clusters}. 
The set of edges of $\Gc$ is denoted by $\Ec$, i.e., $\Gc = \Gc([L], [K], \Ec)$. 

We assume OFDM modulation and assume that the channel in the time-frequency domain follows the
standard block-fading model adopted in a very large number of papers (e.g., see \cite{marzetta2010noncooperative,9336188,9064545}),  
where the channel vectors from UEs to RUs are random but constant over coherence blocks of 
$T$ signal dimensions in the time-frequency domain, which can be identified here as a RB as already illustrated and discussed in Section \ref{sec:intro}. 

Since all our treatment can be formulated on a per-RB basis, we shall neglect the RB index for the sake of notation simplicity. 
We let $\HH \in \CC^{LM \times K}$ denote the channel matrix between all the $K$ UE antennas and all the $LM$ 
RU antennas on a given RB, formed by $M \times 1$ blocks $\hv_{\ell,k}$ in correspondence of the $M$ antennas of RU $\ell$ 
and UE $k$.  Because of the UL pilot allocation (see later), each RU $\ell$ only estimates the channel vectors of the users in $\Uc_\ell$.

As a genie-aided best-case, we define the {\em ideal partial CSI} regime
where each RU $\ell$ has perfect knowledge of the channel vectors $\hv_{\ell,k}$ for $k \in \Uc_\ell$. 
In this regime, the part of the channel matrix $\HH$ known  at the DU serving cluster $\Cc_k$ 
is denoted by $\HH(\Cc_k)$. This matrix has the same dimensions of $\HH$,  
and contains the channel vectors $\hv_{\ell,j}$ in all the  $(\ell, j)$-th blocks 
of dimension $M \times 1$  such that $\ell \in \Cc_k$ and $j \in \Uc_\ell$, and all-zero blocks
of dimension $M \times 1$  in all other cases. 

\begin{figure}[ht]
		\centerline{\includegraphics[width=8cm,height=6cm]{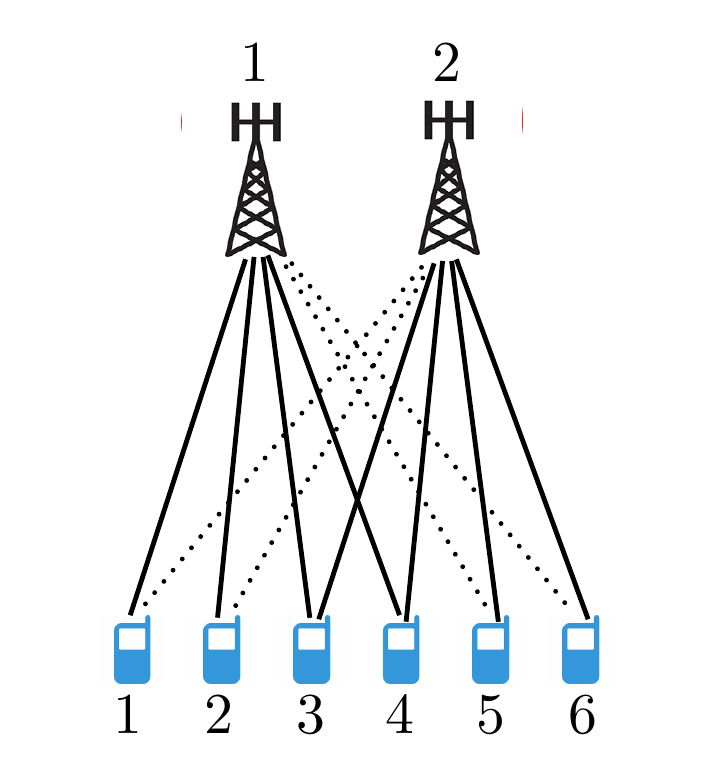}}
		\caption{A simple network with $L = 2$ RUs and $K = 6$ users used as an example. The dotted edges correspond to channels vectors
			that cannot be estimated because of the cluster formation mechanism.}
		\label{clusters2}
\end{figure}

	For example, consider the simple case of $L = 2$ and $K = 6$ as in  Fig.~\ref{clusters2}. 
	Let's focus on user $k = 3$, for which $\Cc_3 = \{1,2\}$. We have $\Uc_1 = \{1,2,3,4\}$ and $\Uc_2 = \{3,4,5,6\}$.
	The complete channel matrix is given by 
	\[ \HH = \left [ \begin{array}{cccccc}
		\hv_{1,1} & \hv_{1,2} & \hv_{1,3} & \hv_{1,4} & \hv_{1,5} & \hv_{1,6} \\
		\hv_{2,1} & \hv_{2,2} & \hv_{2,3} & \hv_{2,4} & \hv_{2,5} & \hv_{2,6} \end{array} \right ] \]
	However, the channel matrix $\HH(\Cc_3)$ is given by  
	\[ \HH(\Cc_3) = \left [ \begin{array}{cccccc}
		\hv_{1,1} & \hv_{1,2} & \hv_{1,3} & \hv_{1,4} & \zerov & \zerov \\
		\zerov & \zerov & \hv_{2,3} & \hv_{2,4} & \hv_{2,5} & \hv_{2,6} \end{array} \right ],  \]
	since users 1 and 2 do not belong to $\Uc_2$ and users 5 and 6 do not belong to $\Uc_1$. Hence, the channels of users that are not associated to a
	given RU cannot be estimated by this RU. 
	
For the individual UE-RU channels, we consider a simplified directional channel model defined as follows.
Let $\Fm$ denote the $M \times M$ unitary DFT matrix with $(m,n)$-elements
$\left[ \Fm \right]_{m,n} = \frac{e^{-j\frac{2\pi}{M} mn}}{\sqrt{M}}$ for  $m, n  = 0,1,\ldots, M-1$. 
Consider the discrete angular support set $\Sc_{\ell,k} \subseteq \{0,\ldots, M-1\}$. We let the channel $\hv_{\ell,k}$ be a random Gaussian vector
in the linear span of the columns of $\Fm$ indexed by $\Sc_{\ell,k}$. In particular, we have
\begin{equation} 
	\hv_{\ell,k} = \sqrt{\frac{\beta_{\ell,k} M}{|\Sc_{\ell,k}|}}  \Fm_{\ell,k} \nuv_{\ell, k},  \label{channel_model}
\end{equation}
where, using a Matlab-like notation, we define $\Fm_{\ell,k} \eqdef \Fm(: , \Sc_{\ell,k})$. Therefore, $\Fm_{\ell,k}$ is a tall unitary matrix
of dimensions $M \times |\Sc_{\ell,k}|$. In particular, for the sake of simplicity we adopt
 the single ring local scattering model (e.g., see \cite{adhikary2013joint}), where $\Sc_{\ell,k}$ is formed by the integer indices 
 $m \in \{0,1,\ldots, M-1\}$ such that the angle $\frac{2\pi m}{M}$ falls (modulo $2\pi$) in the interval 
 $[\theta - \Delta/2, \theta + \Delta/2]$, and where $\theta$ is the angle of the direction between the RU and the UE, and $\Delta$ is the scattering ring angular spread. 
 This model captures the directionality of the channel vectors and it is geometric consistent in the sense that two users with the same direction $\theta$ with respect to a RU will have channels spanning the same subspace. This directionality aspect is particularly relevant for high frequencies (mmWaves) where 
 propagation is dominated by the Line-of-Sight path.
The coefficient $\beta_{\ell,k}$ in (\ref{channel_model}) represents the large scale fading coefficient (LSFC) including distance-dependent 
pathloss, blocking effects, and shadowing. The vector $\nuv_{\ell,k}$ in (\ref{channel_model}) is an $|\Sc_{\ell,k}| \times 1$ i.i.d. Gaussian vector with components 
$\sim \Cc\Nc(0,1)$.  It follows that  $\hv_{\ell,k}$ is a Gaussian zero-mean random vector with covariance matrix 
\begin{equation} 
\Sigmam_{\ell,k} = \frac{\beta_{\ell,k} M}{|\Sc_{\ell,k}|}  \Fm_{\ell,k}   \Fm_{\ell,k} ^\herm. 
\end{equation}

\subsection{Cluster formation} \label{sec:cluster_formation}

We assume that $\tau_p$ signal dimension per RB are dedicated to UL pilots (see \cite{3gpp38211}), and define a codebook of $\tau_p$ orthogonal pilots 
sequences.  The UEs transmit with the same power\footnote{It is customary in communication theory to call ``power'' the average energy per complex symbol. 
We follow here this convention. Notice that the physical power for a symbol rate of $W$ symbols/s is given by $W P^{\rm ue}$ and the physical noise power
(integral of the noise power spectral density over the system bandwidth of $W$ Hz) is given by $W N_0$. 
Hence, the ratio $P^{\rm ue}/N_0$ coincides with the ratio of the physical transmit signal power over the noise power at the receiver output.}
$P^{\rm ue}$,  and we define the system parameter $\SNR \eqdef P^{\rm ue}/N_0$, 
where $N_0$ denotes the noise power spectral density. By the normalization of the channel vectors, the maximum beamforming gain 
averaged over the small scale fading is $\EE[\lVert \frac{M}{|\Sc_{\ell,k}|} \Fm_{\ell,k} \nuv_{\ell, k} \rVert^2] = M$. Therefore, the maximum SNR at the receiver of RU $\ell$ from UE $k$ is 
$\beta_{\ell,k} M \SNR$.  As in \cite{9064545}, we consider that each UE $k$  selects its leading RU $\ell$ as the RU 
with the largest channel gain $\beta_{\ell,k}$ (assumed known) among the RUs with yet a free pilot and satisfying the 
received SNR condition $\beta_{\ell,k} \geq \frac{\eta}{M \SNR}$, where  $\eta > 0$ is a suitable threshold. 
If such RU is not available, then the UE is declared in outage. In our simulations, the UE to leader RU association is performed 
in a greedy manner starting from some UE at random. In practice, users join and leave the system according to some
user activity dynamics, and each new UE joining the system is admitted if it can find a leader RU according to the above conditions. 
After all non-outage UEs $k$ are assigned to their leader RU $\ell = \ell(k)$ and therefore have a pilot index $t = t_k \in [\tau_p]$, the dynamic cluster $\Cc_k$ for each UE $k$  is formed by enrolling successively all RUs $\ell$ 
listed in order of decreasing LSFC for which i) pilot $t_k$ is yet free, ii) the condition $\beta_{\ell,k} \geq \frac{\eta}{M \SNR}$ is satisfied. 
We also consider a maximum cluster size $Q$ such that if more than $Q$ RUs meet the cluster enrollment condition, 
only the $Q$ with the largest LSFC are selected. 
As a result, all UEs $k \in \Uc_\ell$ make use of mutually orthogonal UL pilots. Furthermore, 
$0 \leq |\Uc_\ell| \leq \tau_p$ and $0 \leq |\Cc_k| \leq Q$.

\subsection{Uplink data transmission}

The received $LM \times 1$ symbol vector at the $LM$ RU antennas for a single channel use of the UL is given by
\begin{equation} 
	\yy^{\rm ul} = \sqrt{\SNR} \; \HH \sss^{\rm ul}   + \zz^{\rm ul}, \label{ULchannel}
\end{equation}
where $\sss^{\rm ul} \in \CC^{K \times 1}$ is the vector
of information symbols transmitted by the UEs (zero-mean unit variance and mutually independent random variables) and 
$\zz^{\rm ul}$ is an i.i.d. noise vector with components $\sim \Cc\Nc(0,1)$.  
The goal of cluster $\Cc_k$ is to produce an effective channel observation for symbol $s^{\rm ul}_k$ 
(the $k$-th component of the vector $\sss^{\rm ul}$) from the collectively received signal at the RUs $\ell \in \Cc_k$.  
We  define the receiver {\em unit norm} vector $\vvv_k \in \CC^{LM \times 1}$ formed by $M \times 1$ blocks
$\vv_{\ell,k} : \ell = 1, \ldots, L$, such that $\vv_{\ell,k} = \zerov$ (the identically zero vector) if $\ell \notin \Cc_k$. 
This reflects the fact that only the RUs in $\Cc_k$ are involved in producing a received observation for the detection of user $k$. 
The non-zero blocks $\vv_{\ell,k} :  \ell \in \Cc_k$ contain the receiver combining vectors.
The corresponding scalar combined observation for symbol $s^{\rm ul}_k$ is given by 
\begin{eqnarray}
	r^{\rm ul}_k  & = & \vvv_k^\herm \yy^{\rm ul} \nonumber \\ 
	& =  & \sqrt{\SNR} \vvv_k^\herm \hh_k s^{\rm ul}_k   + \sqrt{\SNR} \vvv_k^\herm \HH_k \sss_k^{\rm ul}  + \vvv_k^\herm \zz^{\rm ul}  \label{received-UL}
\end{eqnarray}
where $\hh_k$ denotes the $k$-th column of $\HH$, 
$\HH_k$ is obtained by deleting the $k$-th column from $\HH$, 
and $\sss_k^{\rm ul} $ is the vector $\sss^{\rm ul}$ after deletion of the $k$-th element. 

For simplicity, we assume that the channel decoder has perfect knowledge of the exact UL SINR value 
\begin{eqnarray} 
	\SINR^{\rm ul}_k 
& = & \frac{  |\vvv_k^\herm \hh_k|^2 }{ \SNR^{-1}  + \sum_{j \neq k} |\vvv_k^\herm \hh_j |^2 },  \label{UL-SINR-unitnorm}
\end{eqnarray}
where $\hh_k$ denotes the $k$-th column of $\HH$. The corresponding UL {\em optimistic ergodic} achievable rate is given by 
\begin{equation}
R_k^{\rm ul} = \EE [ \log ( 1 + \SINR^{\rm ul}_k ) ], \label{ergodic-rate-ul}
\end{equation}
where the expectation is with respect to the small scale fading, while conditioning on the placement of UEs and RU, and on the cluster formation. 

\begin{rem} 
The optimistic ergodic rate is achievable
when the small-scale fading is a stationary ergodic process in the time-frequency domain and a codeword is 
transmitted over a sufficiently large number of small-scale fading states, and somehow the useful signal coefficient and the variance of the interference term in 
\eqref{received-UL} are known to the decoder. These, however, are {\em sufficient conditions} under which the achievability proof of 
\eqref{ergodic-rate-ul} follows easily. In the massive MIMO literature, it is common to consider the UatF
lower bound on \eqref{ergodic-rate-ul}, which contains only the long-term statistics (first and second moments) of the coefficients in $\vvv_k^\herm \hh_j$ in 
\eqref{UL-SINR-unitnorm} and therefore is achievable under less restrictive conditions \cite{marzetta2016fundamentals,9336188}.
It has been shown that unless the useful signal coefficient ``hardens'' such as $|\EE[\vvv_k^\herm \hh_j]|^2$ is large with respect to
Var$(\vvv_k^\herm \hh_k)$, the UatF bound can be very pessimistic \cite{8304782}. This is unfortunately the case for typical layouts of cell-free user-centric 
networks where the channel hardening of very large co-located arrays and rich i.i.d. small-scale fading do not occur.  
In addition, it cannot be excluded that with some
{\em universal decoding scheme} \cite{feder1998universal}  the rate in  \eqref{ergodic-rate-ul} can be achieved or at least closely approached. 
For these reasons, we believe that the optimistic ergodic rate reflects more accurately the actual achievable performance of the system at hand, than the 
overly conservative UatF lower bound.
\hfill $\lozenge$
\end{rem}

\subsection{Downlink data transmission} 

The signal corresponding to one channel use of the DL at the receiver of UE $k$ is given by 
\begin{equation} 
	y_k^{\rm dl} = \hh_k^\herm \xx  + z_k^{\rm dl},  \label{DLchannel}
\end{equation}
where the transmitted vector $\xx \in \CC^{LM \times 1}$ is formed by all the signal samples sent collectively from the RUs. Without loss of generality
we can incorporate a common factor ${\SNR}^{-1/2}$ in the LSFCs, which is equivalent to rescaling the noise at the UE receivers such that  
$z_k^{\rm dl} \sim \Cc\Nc(0, \SNR^{-1})$, while keeping $\beta_{\ell,k}$ for all $(\ell,k)$ identical to the UL case.
Let $\sss^{\rm dl} \in \CC^{K \times 1}$ denote the vector of information bearing symbols for the $K$ users, assumed to 
be zero mean, independent, with variance $q_k \geq 0$. 
Under a general linear precoding scheme, we have
\begin{equation} 
	\xx = \UU \sss^{\rm dl}, \label{linear-precoding}
\end{equation}
where $\UU \in \CC^{LM \times K}$ is the overall precoding matrix, formed by $M \times 1$ blocks $\uv_{\ell,k}$ such that
$\uv_{\ell,k} = \zerov$ if $\ell \notin \Cc_k$.  The non-zero blocks $\uv_{\ell,k} :  \ell \in \Cc_k$ contain the precoding vectors. Using (\ref{linear-precoding}) in (\ref{DLchannel}), we have
\begin{eqnarray}
	y^{\rm dl}_k & = & \hh_k^\herm \uu_k s^{\rm dl}_k   + \sum_{j \neq k} \hh_k^\herm \uu_j s^{\rm dl}_j  + z^{\rm dl}_k, 
\end{eqnarray}
where $\uu_k$ is the $k$-th column of $\UU$.  The resulting DL (optimistic) SINR is given by 
\begin{eqnarray}
	\SINR^{\rm dl}_k & = & \frac{|\hh_k^\herm \uu_k|^2 q_k}{\SNR^{-1} + \sum_{j\neq k}   |\hh_k^\herm \uu_j|^2 q_j }, \label{DL-SINR} 
\end{eqnarray}
where $q_k$ is the DL transmit (Tx) power for the data stream to UE $k$. As for the UL, we consider the  DL optimistic ergodic rate given by 
\begin{equation}
	R_k^{\rm dl} = \EE [ \log ( 1 + \SINR^{\rm dl}_k ) ]. \label{ergodic-rate-dl}
\end{equation}
Assuming that the columns of the precoder $\UU$ have unit norm, we have that the total DL Tx power collectively transmitted by the RUs is given by 
\begin{align} 
	P^{\rm dl}_{\rm tot} &= \trace \left ( \EE [ \xx \xx^\herm ] \right ) = \trace \left ( \UU \diag(q_k : k \in [K]) \UU^\herm \right ) \\
	&= \trace \left ( \UU^\herm \UU \diag(q_k : k \in [K]) \right ) = \sum_{k=1}^K q_k.  
\end{align}
We assume the UL and DL total transmit power to be balanced. This imposes the condition $\sum_{k=1}^K q_k = K$. 

\section{Uplink receive schemes} \label{sec:ul_perfect_csi}

We illustrate the UL linear receive schemes for the case of ideal partial CSI. In our simulations, we shall use
the CSI estimated as described in Section \ref{CSI-est} and simply plug the estimated channel vectors in place of the ideally known channel vectors.

In this work we consider two UL receive schemes referred to as cluster-level zero-forcing (CLZF) and 
local linear MMSE (LMMSE) with cluster-level combining \cite{goettsch2021impact}. The schemes are described in the following. 

\subsection{Cluster-Level Zero-Forcing (CLZF)}  \label{gzf_combining}

For a given UE $k$ with cluster $\Cc_k$,  we define the set $\Uc(\Cc_k) \eqdef \bigcup_{\ell \in \Cc_k} \Uc_\ell$ of UEs served by at least one RU in $\Cc_k$. 
Let $\hh_k(\Cc_k)$ denote the $k$-th column of $\HH(\Cc_k)$ and let $\HH_k(\Cc_k)$ denote the residual matrix after deleting the $k$-th column. 
The CLZF receiver vector is obtained as follows.  Let $\overline{\hh}_k(\Cc_k) \in \CC^{|\Cc_k|M \times 1}$ and $\overline{\HH}_k(\Cc_k) \in  \CC^{|\Cc_k|M \times (|\Uc(\Cc_k)|-1)}$ the vector and matrix
obtained from $\hh_k(\Cc_k)$ and $\HH_k(\Cc_k)$, respectively, after removing all the $M$-blocks of rows corresponding to 
	RUs $\ell \notin \Cc_k$ and all the (all-zero) columns corresponding to UEs $k' \notin \Uc(\Cc_k)$. Consider the 
	singular value decomposition (SVD) 
	\begin{equation}
		\overline{\HH}_k(\Cc_k) = \overline{\AA}_k \overline{\SSS}_k \overline{\BB}_k^\herm, 
	\end{equation}
	where the columns of the tall unitary matrix $\overline{\AA}_k$ form an orthonormal basis for the column span of $\overline{\HH}_k(\Cc_k)$, such that 
	the orthogonal projector onto the orthogonal complement of the interference subspace is given by 
$\overline{\PP}_k = \Id - \overline{\AA}_k \overline{\AA}_k^\herm$, 
and define the unit-norm vector 
	\begin{equation} 
	\overline{\vvv}_k = \overline{\PP}_k \overline{\hh}_k(\Cc_k) / \| \overline{\PP}_k \overline{\hh}_k(\Cc_k) \|. 
	\end{equation} 
	Hence, the CLZF receiver vector $\vvv_k$ is given by expanding $\overline{\vvv}_k$ by reintroducing the missing blocks of 
	all-zero $M \times 1$ vectors $\zerov$ in correspondence of the RUs $\ell \notin \Cc_k$. 
	
	Because of the channel correlation model, it may happen that there exists some UE $j \in \Uc(\Cc_k), j \neq k$, such that 
	the $k$-th and $j$-th column of $\HH(\Cc_k)$ are co-linear. In this case, this column is extracted from the channel matrix and it is not included in the CLZF computation. In practice, this means that the interference on user $k$ caused by such user $j$ is taken in directly, without mitigation by linear projection.

	\subsection{Local LMMSE with cluster-level combining}

	In this case, each RU $\ell$ makes use of locally computed receiving vectors $\vv_{\ell,k}$ for 
	its users $k \in \Uc_\ell$. Let $\yv_\ell^{\rm ul}$ denote
	the $M \times 1$ block of $\yy^{\rm ul}$ corresponding to RU $\ell$. For each $k \in \Uc_\ell$, RU $\ell$ computes locally
	\begin{eqnarray} 
		r^{\rm ul}_{\ell,k} = \vv_{\ell,k}^\herm \yv_\ell^{\rm ul}.  \label{suca}
	\end{eqnarray}
The symbols $\{r^{\rm ul}_{\ell,k} : k \in \Uc_\ell\}$ are sent to the DU serving cluster $\Cc_k$, which computes the cluster-level combined symbol 
\begin{equation} 
r^{\rm ul}_k = \sum_{\ell \in \Cc_k} w^*_{\ell,k} r^{\rm ul}_{\ell,k} = \wv^\herm_k \rv^{\rm ul}_k,   \label{combining}
\end{equation}
where $w_{\ell,k}$ is the combining coefficient of RU $\ell$ for UE $k$, and $\wv_k$ and $\rv_k$ are vectors formed by stacking $w_{\ell,k}$ and $r^{\rm ul}_{\ell,k}$ of all RUs $\ell \in \Cc_k$, respectively.
	
	A classical and effective choice for the receiver vector $\vv_{\ell,k}$  is based on LMMSE estimation. 
	In this case, we distinguish between the known part of the interference, 
	i.e., the term  $\sum_{j \in \Uc_\ell : j \neq k}  \hv_{\ell,j} s_j^{\rm ul}$, and the unknown part of the interference, 
	i.e., the term  $\sum_{j \notin \Uc_\ell} \hv_{\ell,j} s_j^{\rm ul}$ in $\yv_\ell^{\rm ul}$. 
	The receiver treats the unknown part of the interference plus noise as a white vector with known variance per component. 
	The covariance matrix of this term is given by 
	\begin{equation} 
		\Xim_\ell = \EE \left [ \left ( \sqrt{\SNR} \sum_{j \notin \Uc_\ell} \hv_{\ell,j} s_j^{\rm ul}  + \zv_\ell^{\rm ul} \right )
		\left ( \sqrt{\SNR} \sum_{j \notin \Uc_\ell} \hv_{\ell,j} s_j^{\rm ul}  + \zv_\ell^{\rm ul} \right )^\herm \right ] \nonumber \\ 
		\hspace{-3.37cm} = \Id +  \sum_{j \notin \Uc_\ell} \frac{\beta_{\ell,j} M \SNR }{|\Sc_{\ell,j}|} \Fm_{\ell,j} \Fm_{\ell,j}^\herm, 
	\end{equation} 
	where $\zv_\ell^{\rm ul} \sim \Cc\Nc(0,1)$ is AWGN at RU $\ell$.
	Taking the trace and dividing by $M$ we find the equivalent variance per component
	\begin{equation} 
		\sigma^2_\ell = \frac{1}{M} \trace ( \Xim_\ell ) =   1 + \SNR \left ( \sum_{j \neq \Uc_\ell}  \beta_{\ell,j} \right ).  \label{sigmaell}
	\end{equation}
	Under this assumption, we have that the LMMSE receiving vector is given by 
	\vspace{-.15cm}
	\begin{equation} 
		\vv_{\ell,k} = \left ( \sigma_\ell^2 \Id + \SNR \sum_{j \in \Uc_\ell} \hv_{\ell,j} \hv_{\ell,j}^\herm \right )^{-1} \hv_{\ell,k}.  \label{eq:lmmse}
	\end{equation}
	For the cluster-level combining coefficients, we consider the maximization of the SINR after combining.  
	The effective received signal model at RU $\ell \in \Cc_k$ relative to UE $k$ can be written as
	\begin{equation} 
		r^{\rm ul}_{\ell,k} = \sqrt{\SNR} \left ( g_{\ell, k,k} s^{\rm ul}_k + \sum_{j \in \Uc_\ell: j \neq k} g_{\ell, k,j} s^{\rm ul}_j  \right ) +  \vv_{\ell, k}^\herm \xiv_\ell  \label{effective model} 
	\end{equation} 
	where we define
		$g_{\ell,k,j} = \vv_{\ell,k}^\herm \hv_{\ell,j}$
	and let $\xiv_\ell$ the unknown interference plus noise vector, assumed $\sim \Cc\Nc(\zerov, \sigma_\ell^2 \Id)$. 
	Stacking $\{r^{\rm ul}_{\ell,k} : \ell \in \Cc_k\}$ as a $|\Cc_k| \times 1$ column vector $\rv^{\rm ul}_k$, we can write the output symbols of cluster $\Cc_k$ relative to UE $k$ as
	\begin{equation}
		\rv^{\rm ul}_k = \sqrt{\SNR}  \left ( \av_k s^{\rm ul}_k  +  \Gm_k \sv_k^{\rm ul} \right ) + \zetav_k,  \label{cluster-level-ch-model}
	\end{equation} 
	where 
		$\zetav_k = \{ \vv_{\ell, k}^\herm \xiv_\ell  : \ell \in \Cc_k \}$
	has the covariance matrix given by 
		$\Dm_k = \diag \left \{ \sigma_\ell^2 \|\vv_{\ell,k}\|^2 : \ell \in \Cc_k \right \}$
	and $\av_k = \{ g_{\ell, k,k} : \ell \in \Cc_k\}$.

	The matrix $\Gm_k \in \CC^{|\Cc_k| \times (|\Uc(\Cc_k)| - 1)}$ contains elements $g_{\ell, k, j}$ in position corresponding to 
	RU $\ell \in \Cc_k$ and UE $j \in \Uc(\Cc_k)\setminus k$ (after a suitable index reordering) if $(\ell,j) \in \Ec$, and zero elsewhere. 
	The vector $\sv_k^{\rm ul} \in \CC^{(|\Uc(\Cc_k)| - 1) \times 1}$ contains the symbols of all users $j \in \Uc(\Cc_k) \setminus k$. 
	Then, the total interference plus noise covariance matrix given the available CSI is
	\begin{equation}
		\Gammam_k = \Dm_k  + \SNR \; \Gm_k \Gm^\herm_k.  \label{eq:interference_plus_noise_cov}
	\end{equation}
The SINR for user $k$ resulting from the channel model \eqref{cluster-level-ch-model} with combining is given by 
	\begin{equation} 
		\SINR^{\rm cl}_k = \frac{\SNR \; \wv_k^\herm \av_k \av_k^\herm \wv_k}{\wv_k^\herm \Gammam_k  \wv_k}.  \label{SINRnom}
	\end{equation}
	The maximization of \eqref{SINRnom} with respect to $\wv_k$ amounts to 
	find the maximum generalized eigenvalue of the matrix pencil 
	$(\av_k \av_k^\herm, \Gammam_k )$. Since the matrix $\av_k \av_k^\herm$ has rank 1 and therefore it has only one non-zero eigenvalue, 
	the solution is readily given by 
	\begin{equation}
		\wv_k = \Gammam_k^{-1} \av_k, \label{eq:w_opt}
	\end{equation}
yielding the SINR 
	\begin{equation} 
		\SINR^{\rm cl}_k = \SNR \; \av_k^\herm \Gammam_k^{-1} \av_k. \label{SINRnom1}
	\end{equation}
For the LMMSE with cluster-level combining scheme, the overall received vector is obtained by forming the vector $\overline{\vvv}_k$ by stacking the $|\Cc_k|$ blocks  of dimensions $M \times 1$ given by  $w_{\ell,k} \vv_{\ell,k}$ on top of each other and normalizing such that $\overline{\vvv}_k$ has unit norm. 
After expanding $\overline{\vvv}_k$ to $\vvv_k$ of dimension $LM \times 1$ by inserting the all-zero blocks corresponding to the RUs $\ell \notin \Cc_k$, the resulting  SINR is again given by (\ref{UL-SINR-unitnorm}) and does not coincide in general with \eqref{SINRnom1} since it takes into account the true interference from 
all users, and not only the known part due to the partial CSI. Nevertheless, the SINR in \eqref{SINRnom} and its maximization leading to 
\eqref{SINRnom1} represent the best guess given the available CSI and the statistical information on the overall additional interference plus noise power. 

It is interesting to notice that the scheme differs from the distributed LSFD in \cite{9336188}, as the LSFD relies on the expected products of
the combining and channel vectors. In contrast, we use the instantaneous channel realization estimate and the instantaneous receive vector for the computation of the
combining coefficients.
We shall provide a more detailed comment on the difference between the two schemes in Section \ref{sec:other-schemes}.

\section{UL-DL duality and Downlink precoding schemes} \label{sec:dl_schemes}

We shall consider expressions of {\em nominal}, i.e., assumed, SINR, where the channels $\hv_{\ell,k}$ with $(\ell,k) \in \Ec$ are known, while for $(\ell,k) \notin \Ec$, only the LSFCs of their channels are known, and the channel vector spatial distribution is assumed isotropic. This treatment differs from what done in 
\cite{9064545} where it is assumed that, if RU $\ell$ serves UE $j$ but not UE $k$, $\EE[ |  \hh^\herm_k \vvv_j |^2 ]$ can still be estimated accurately 
at RU $\ell$, where $\hv_{\ell,k}^\herm \vv_{\ell,j} $ is a non-zero component of $\hh^\herm_k \vvv_j$. 

\subsection{UL-DL SINR duality for partially known channels}

Consider the UL SINR given in (\ref{UL-SINR-unitnorm}). By construction we have that $\vv_{\ell,k} = \zerov$ for all $\ell \notin \Cc_k$. Therefore, the term at the numerator
$ \theta_{k,k} = \big | \vvv_k^\herm \hh_k \big |^2 = \big| \sum_{\ell\in \Cc_k} \vv_{\ell,k}^\herm \hv_{\ell,k} \big|^2 $
contains only known channels and it can thus be used in the nominal SINR expression. 
The terms at the denominator take on the form
\begin{eqnarray}
	\theta_{j,k} & = & \left | \vvv_k^\herm \hh_j \right |^2 = \bigg | \sum_{\ell \in \Cc_k} \vv_{\ell,k}^\herm \hv_{\ell,j} \bigg |^2 \\
	& = & \bigg| \sum_{\ell \in \Cc_k \cap \Cc_j} \vv_{\ell,k}^\herm \hv_{\ell,j}  + \sum_{\ell \in \Cc_k \setminus \Cc_j} \vv_{\ell,k}^\herm \hv_{\ell,j} \bigg|^2 \label{known-unknown}
\end{eqnarray}
where for $\ell \in \Cc_k \cap \Cc_j$ the channel $\hv_{\ell,j}$ is known, while for $\ell \in \Cc_k \setminus \Cc_j$ the channel $\hv_{\ell,j}$ is not known. 
Taking the conditional expectation of the term in (\ref{known-unknown}) given all the known CSI $\Ec_k$ at $\Cc_k$, and noticing that the channels for different
$(\ell,j)$ pairs are independent and have mean zero, we find
\begin{align} 
	&\EE[ \theta_{j,k} | \Ec_k] \nonumber \\
	&= \bigg | \sum\limits_{\ell \in \Cc_k \cap \Cc_j} \vv_{\ell,k}^\herm \hv_{\ell,j} \bigg |^2 + 	\sum\limits_{\ell \in \Cc_k\setminus \Cc_j} \frac{\beta_{\ell,j} M}{|\Sc_{\ell,j}|} \vv_{\ell,k}^\herm \Fm_{\ell,j} \Fm_{\ell,j}^\herm \vv_{\ell,k}.  \nonumber
\end{align}
Finally, under the isotropic assumption, we replace the actual covariance matrix of the unknown channels with
a scaled identity matrix with the same trace, i.e., we have
\begin{equation} 
	\EE[ \theta_{j,k} | \Ec_k] \approx \bigg | \sum_{\ell \in \Cc_k \cap \Cc_j} \vv_{\ell,k}^\herm \hv_{\ell,j} \bigg |^2 + 
	\sum_{\ell \in \Cc_k\setminus \Cc_j} \beta_{\ell,j} \| \vv_{\ell,k} \|^2.  \nonumber
\end{equation}
Using the fact that $\vvv_k$ is a unit-norm vector and assuming that the  $M \times 1$ blocks $\vv_{\ell,k}$ have the same norm, we can further approximate
$\| \vv_{\ell,k}\|^2 \approx \frac{1}{|\Cc_k|}$. Therefore, the resulting {\em nominal} UL SINR is given by 
\begin{align} 
&\SINR_k^{\rm ul-nom} \nonumber \\
	& =  \frac{\left | \vvv_k^\herm \hh_k \right |^2}{\SNR^{-1} + \sum\limits_{j \neq k}  \Big( 
		\Big | \sum\limits_{\ell \in \Cc_k \cap \Cc_j} \vv_{\ell,k}^\herm \hv_{\ell,j} \Big |^2 + \frac{1}{|\Cc_k|} \sum\limits_{\ell \in \Cc_k\setminus \Cc_j} \beta_{\ell,j}  \Big)} \nonumber \\
	& =  \frac{\theta_{k,k}}{\SNR^{-1} + \sum_{j\neq k} \widetilde{\theta}_{j,k} },   \label{UL-nom-SINR}
\end{align}
where define $\theta_{k,k} = | \vvv_k^\herm \hh_k |^2$  and 
for $j \neq k$ as
\begin{equation} 
	\widetilde{\theta}_{j,k} =  
	\bigg | \sum_{\ell \in \Cc_k \cap \Cc_j} \vv_{\ell,k}^\herm \hv_{\ell,j} \bigg |^2 + \frac{1}{|\Cc_k|} \sum_{\ell \in \Cc_k\setminus \Cc_j} \beta_{\ell,j}.  \label{new-theta-jk}
\end{equation}
Notice that the term at the denominator of the nominal UL SINR given by $\frac{1}{|\Cc_k|}  \sum_{j \neq k} \sum_{\ell \in \Cc_k\setminus \Cc_j} \beta_{\ell,j} $
is the contribution of the interference power per RU caused by other UEs $j \neq k$ to the RUs in $\Cc_k$, but not also in $\Cc_j$.
We refer to \eqref{UL-nom-SINR} as the ``nominal'' UL SINR since this is the SINR that the processor of cluster $\Cc_k$ can estimate from its CSI knowledge. 

Next, we consider the DL SINR given in (\ref{DL-SINR}) under the assumption that the DL precoding vectors are identical to the 
UL receive vectors, i.e., $\uu_k = \vvv_k$ for all $k \in [K]$. 
The numerator takes on the form $\theta_{k,k} q_k$ where $\theta_{k,k}$ is the same as before and contains all known channels. 
Focusing on the terms at the denominator and taking into account that the vectors $\uu_j$ are non-zero only for the $M\times 1$ blocks corresponding to RUs $\ell \in \Cc_j$, we have
\begin{eqnarray}
	\theta_{k,j} & = & \left | \hh_k^\herm \uu_j \right |^2 = \bigg | \sum_{\ell \in \Cc_j} \hv_{\ell,k}^\herm \uv_{\ell,j} \bigg |^2 \\
	& = & \bigg | \sum_{\ell \in \Cc_j \cap \Cc_k} \hv_{\ell,k}^\herm  \uv_{\ell,j}  + \sum_{\ell \in \Cc_j \setminus \Cc_k} \hv_{\ell,k}^\herm \uv_{\ell,j} \bigg |^2, \label{known-unknown-DL}
\end{eqnarray}
where, as before,  for $\ell \in \Cc_j \cap \Cc_k$ the channel $\hv_{\ell,k}$ is known, while for $\ell \in \Cc_j \setminus \Cc_k$ the channel $\hv_{\ell,k}$ 
is not known.  Taking the conditional expectation of the term in (\ref{known-unknown-DL}) given all the known channel state information, 
we find
\begin{align} 
	&\EE[ \theta_{k,j} | \Ec_k] \nonumber \\
	&= \bigg | \sum\limits_{\ell \in \Cc_j \cap \Cc_k} \hv_{\ell,k}^\herm \uv_{\ell,j} \bigg |^2 + 
	\sum\limits_{\ell \in \Cc_j \setminus \Cc_k} \frac{\beta_{\ell,k} M}{|\Sc_{\ell,k}|} \uv_{\ell,j}^\herm \Fm_{\ell,k} \Fm_{\ell,k}^\herm \uv_{\ell,j}. \nonumber 
\end{align}
Using again the isotropic assumption, we replace the actual covariance matrix of the unknown channels with
a scaled identity matrix with the same trace, i.e., we have
\begingroup
\allowdisplaybreaks
\begin{eqnarray} 
	\EE[ \theta_{k,j} | \Ec_k] & \approx & 
	\bigg | \sum_{\ell \in \Cc_j \cap \Cc_k} \hv_{\ell,k}^\herm \uv_{\ell,j}  \bigg |^2 + 
	\sum_{\ell \in \Cc_j\setminus \Cc_k} \beta_{\ell,k} \| \uv_{\ell,j} \|^2 \nonumber \\
	& \approx & 
	\bigg | \sum_{\ell \in \Cc_j \cap \Cc_k} \hv_{\ell,k}^\herm \uv_{\ell,j}  \bigg |^2 + 
	\frac{1}{|\Cc_j|} \sum_{\ell \in \Cc_j\setminus \Cc_k} \beta_{\ell,k}. \nonumber
\end{eqnarray}
\endgroup
Therefore, the resulting nominal DL SINR is given by 
\begin{align} 
&\SINR_k^{\rm dl-nom} \nonumber \\
	& = 
	\frac{\left | \hh_k^\herm \uu_k  \right |^2 q_k}{\SNR^{-1} + \sum\limits_{j\neq k} \Biggl( 
			\Big | \sum\limits_{\ell \in \Cc_j \cap \Cc_k} \hv_{\ell,k}^\herm \uv_{\ell,j} \Big |^2 + \frac{1}{|\Cc_j|} \sum\limits_{\ell \in \Cc_j\setminus \Cc_k} \beta_{\ell,k} \Biggr) q_j }  \nonumber \\
	& = \frac{\theta_{k,k} q_k}{\SNR^{-1} + \sum_{j\neq k} \widetilde{\theta}_{k,j} q_j}, \label{DL-nom-SINR} 
\end{align}
where 
\begin{equation} 
	\widetilde{\theta}_{k,j} = 
	\bigg | \sum_{\ell \in \Cc_j \cap \Cc_k} \hv_{\ell,k}^\herm \uv_{\ell,j} \bigg |^2 + \frac{1}{|\Cc_j|} \sum_{\ell \in \Cc_j\setminus \Cc_k} \beta_{\ell,k}  .
	\label{new-theta-kj}
\end{equation}
Given the symmetry of the coefficients, an UL-DL duality exists for the nominal SINRs. This can be used to calculate the DL Tx power allocation $\{q_k : k \in [K]\}$ that achieves DL nominal SINRs equal to the UL nominal SINRs with uniform UL Tx power per UE. 

\begin{rem} \label{balanced-power}
We focus here on the case of balanced UL-DL SINRs for the following reasons: 1) from a theoretical viewpoint, it is easier and more direct to illustrate the SINR duality in this case, and then generalize it to unbalanced (but proportional) SINRs as done later on in Section \ref{unbalanced-power}; 2) from a practical viewpoint, 
in a cell-free architecture the cost, size, and scale manufacturing of RUs plays a very important role for these systems to be attractive and effectively deployed. 
Hence, it is likely to assume that each ``antenna element'' 
(including physical antenna, low-noise amplifier (LNA)/power amplifier (PA) up-down conversion and A/D and D/A conversion)
of a multiantenna RU has (roughly) the same characteristics of the corresponding element in a UE. In fact, there exist already proposals to reuse 
UE chipsets to create scalable and economically viable RUs. Therefore, the balanced total transmit power when the total number of users and the total number of 
RU antenna elements is similar is a meaningful working assumption;  3) achieving balanced UL-DL SINRs has nothing to do with the fact that, typically, the traffic load in the DL is much larger than in the UL. In fact, given balanced UL-DL SINRs, the different traffic load can be matched by scheduling, i.e., allocating a different number of 
transmission resources to the UL and DL, respectively. The allocation of  transmission resources (i.e., time-frequency RBs) to UL and DL is a much more effective way than ``power allocation'' to match the traffic load requests, since it acts on the pre-log factor of the time-averaged rate (throughput) rather than on the term inside the logarithm; 
4) working with balanced SINRs has the non-trivial advantage that only the UL receiving vectors must be calculated for each new CSI pilot round, while the DL precoding vectors are automatically obtained as a byproduct.
\hfill $\lozenge$
\end{rem}

In particular, we choose the target DL SINRs $\{\gamma_k\} \eqdef \{\SINR_k^{\rm ul-nom}\}$ for all $k$. 
The system of (non-linear) equations in the power allocation vector $\qv = \{q_k\}$ given by 
\[ \SINR^{\rm dl-nom}_k = \gamma_k, \;\;\; \forall \; k = 1\ldots, K \]
can be rewritten in the more convenient linear form (see \cite{Viswanath-Tse-TIT03})
\begin{equation} 
	\left ( \Id  - \diag(\muv) \Thetam \right ) \qv  = \frac{1}{\SNR} \muv,  
\end{equation} 
by defining the vector $\muv$ with elements 
\begin{equation}
	\mu_k = \frac{\gamma_k}{(1 + \gamma_k) \theta_{k,k}} \label{mumu}
\end{equation}
and the matrix $\Thetam$ with $(k,j)$ elements $\theta_{k,k}$ on the diagonal and $\widetilde{\theta}_{k,j}$ in the off-diagonal positions. 
Since the target (nominal) SINRs $\{\gamma_k\}$ are achievable, it can be shown that  the above system of equations has a non-negative solution, given by 
\begin{equation} 
	\qv^\star  = \frac{1}{\SNR} \left ( \Id  - \diag(\muv) \Thetam \right )^{-1}  \muv. \label{eq_q_duality}
\end{equation} 
In addition, it is immediate to show that this solution satisfies the total power constraint $\sum_{k=1}^K q^\star_k = K$, i.e., the UL and DL have balanced total power. 
This is shown explicitly as follows. 
Since in the UL case the transmit symbol energies are all equal to 1, for the choice of target SINRs $\gamma_k = \SINR_k^{\rm ul-nom}$
 the following equation must hold:
\[  \onev = \frac{1}{\SNR} \left ( \Id  - \diag(\muv) \Thetam^\transp \right )^{-1}  \muv \]
Hence, it follows that
\begin{eqnarray}
	K & = & \onev^\transp \onev \nonumber \\
	& = & \frac{1}{\SNR} \onev^\transp \left ( \Id  - \diag(\muv) \Thetam^\transp \right )^{-1}  \muv \nonumber \\
	& = & \frac{1}{\SNR} \onev^\transp \left (\diag(1/\mu_1, \ldots, 1/\mu_K)  - \Thetam^\transp \right )^{-1} \onev \nonumber \\
	& = & \frac{1}{\SNR} \onev^\transp \left (\diag(1/\mu_1, \ldots, 1/\mu_K)  - \Thetam \right )^{-1} \onev \nonumber \\
	& = & \frac{1}{\SNR} \onev^\transp \left (\Id  -  \diag(\muv) \Thetam \right )^{-1} \muv \nonumber \\
	& = & \sum_{k=1}^K q^\star. 
\end{eqnarray}
In conclusion, we propose to use as DL precoders the same vectors already computed as UL receivers/combiners. 
This has the advantage of achieving balanced UL and DL (nominal) SINRs as well as total transmit power, and most importantly, that 
no additional DL precoding computation is required. 
In particular, in this work we consider CLZF precoding and LMMSE with cluster-level combining precoding. 

\subsection{The case of unbalanced UL and DL}  \label{unbalanced-power}

First of all, we should notice that (in line with the observation made in Remark \ref{balanced-power}), the user rates expressed by 
\eqref{ergodic-rate-ul} have little to do with the actual user throughputs, defined as the long-term averaged rate over a sequence of RBs. 
In  general, the total number of users in the system $K_{\rm tot}$ may be much larger than the number of users $K$ simultaneously active on a single RB, which is
the quantity considered here. On top of the PHY described in this paper, there exist several other mechanisms such as resource allocation and fairness scheduling, 
such that active users need not be transmitting/receiving on each RB. Considering a long sequence of slots (e.g., formed by blocks of RBs) indexed by a slot time $s$, 
the throughput of user $k \in [K_{\rm tot}]$ is given by 
\begin{equation}
 \overline{R}_k = \lim_{S \rightarrow \infty} \frac{1}{S} \sum_{s=1}^S R_k[s] \label{throughput}
 \end{equation}
where $R_k[s]$ coincides with \eqref{ergodic-rate-ul} on the slots $s$ where user $k$ is scheduled, and it is equal to 0 on the slots $s$ where user $k$ is not 
scheduled. Under mild conditions of stationarity and ergodicity of the scheduling policy and of the fading processes, 
the throughput in \eqref{throughput} converges to the average quantity
\begin{equation} 
\overline{R}_k = \alpha_k \EE[ R_k ] \label{throughput1}
\end{equation}
where $\alpha_k$ is the activity fraction of user $k$, i.e., the fraction of slots where user $k$ is scheduled (active), and 
$\EE[ R_k ]$ is the rate in  \eqref{ergodic-rate-ul}, further averaged over the active user subset of $K - 1$ simultaneously scheduled users out of 
the $K_{\rm tot}$, active together with user $k$. 

Hence, a way to match the UL and DL different traffic demand consists of scheduling a different number of UL and DL slots per TDD frame. 
Also, a way to achieve throughput fairness consists of using a fairness oriented scheduling policy, e.g., following the general theory of 
network utility function maximization (e.g., see \cite{neely2010stochastic}), as applied for example in \cite{shirani2010mimo} in the MU-MIMO case. 
The extension of this type of fairness scheduling approaches to cell-free user-centric networks in a scalable way, i.e., 
without the need of a fully centralized scheduler, is generally a very interesting problem for further research. 

However, it may happen that for some technology reason the total transmit power in the DL is significantly different from the total transmit power in the UL. 
In this case, it does not make sense to reduce artificially the transmit power of the strongest one to match the weakest. Hence, in the rest of this section
we provide the (simple) extension of the UL-DL duality provided before, to the unbalanced total transmit power case. 

Let $P^{\rm ru}$ denote the average RU transmit power such that the total transmit power in  the DL is $L P^{\rm ru}$. 
We consider a virtual UL with per-UE transmit power given by $(L/K) P^{\rm ru}$ and the corresponding virtual UL SNR parameter 
$\SNR^{\rm u} = \frac{L P^{\rm ru}}{KN_0}$. For such virtual UL with receivers $\{ \vvv_k\}$, the nominal SINR is obtained by replacing 
$\SNR$ by $\SNR^{\rm ul}$  in (\ref{UL-nom-SINR}). 
Notice that the vectors $\{\vvv_k\}$ may or may not correspond to the actual receiving vectors for the actual UL. In particular, for the CLZF scheme, 
the receiving vectors are independent of SNR, therefore they are the same for the actual and virtual UL. 
In contrast, with the LMMSE scheme, the receiving vectors depend on SNR and therefore they need to be recomputed for the virtual UL. 
In any case, using $\uu_k = \vvv_k$ for all $k \in [K]$ for the (actual) DL, and 
obtaining the DL normalized transmit powers according to (\ref{eq_q_duality}) with the substitution of 
$\SNR$ by $\SNR^{\rm ul}$, we obtain a set of DL transmit power factors of the $K$ DL streams such that 
1) the same nominal SINRs of the virtual UL are achieved, and 2) the total average DL transmit power is $L P^{\rm ru}$ as desired. 

Of course, in general a combined approach to unbalanced UL and DL traffic an be implemented, where we use the virtual UL approach to take into account the
different total transmit available power, and yet use scheduling to adjust the throughput to the individual user demands and fairness.

\section{Comparison with other proposed UL and DL schemes}  \label{sec:other-schemes}

Several UL receive/combining and DL precoding schemes have been proposed in the literature. Here we review a few. 

For the UL,  the local linear MMSE processing in combination with LSFD proposed in \cite{8845768} and \cite{demir2021cellfree} is similar to 
our  local LMMSE with cluster-level combining introduced before. The difference is in the computation of the weighting coefficients, that with LSFD are dependent on the expected products of the combining and channel vectors $\{\EE[g_{\ell,k,j}]  : (\ell, k) \in \Ec \}$, and $\EE[\Gm_k \Gm^\herm_k]$. In contrast, the 
scheme in this paper uses the instantaneous channel realization for the computation of the combining coefficients. 
The received signal model at RU $\ell$ for UE $k$ is given again by (\ref{suca}).  The local estimates are sent to the DU processing cluster $\Cc_k$ 
and combined according to (\ref{combining}). However, in LSFD the weights are computed in a different way, i.e., 
\begin{equation}
	\wv_k = (\Gammam_k^\text{LSFD})^{-1} \EE[\av_k],
\end{equation}
where 
\begin{equation}
	\Gammam_k^\text{LSFD} = \Dm_k  + \SNR \; \sum_{j \neq k: j \in \Uc_\ell} \EE[\Gm_k(:,j) \Gm_k(:,j)^\herm].
\end{equation}
Recall that the matrix $\Gm_k$ of dimension $|\Cc_k| \times ( | \Uc(\Cc_k)| - 1)$ contains elements $g_{\ell, k, j}$ in position corresponding to 
RU $\ell$ and UE $j$ if $(\ell,j) \in \Ec$, and zero elsewhere. This results in
$\EE[\Gm_k(:,j) \Gm_k(:,j)^\herm]_{m,n}$ given by 
\begin{equation}
	 {\begin{cases}
			\EE[ \vv_{m,k}^\herm \hv_{m,j}  (\vv_{n,k}^\herm \hv_{n,j})^\herm   ] ,& \text{if $(m, k) \in \Ec $, $(n, k) \in \Ec $}, j \in \Uc(\Cc_k) \setminus k \\ 
			0 ,& \textup{otherwise,} \end{cases}}  \label{lsfd_matrix_G}
\end{equation}
and 
\begin{equation}
	\EE[\av_k] = \left\{ \EE[g_{\ell,k,k}] : \ell \in \Cc_k \right\}.
\end{equation}
Instead of the instantaneous channel and combining vectors taking into account small-scale fading as in (\ref{eq:w_opt}), 
the expectation (based on large-scale fading) is used. The choice of these combining coefficients is motivated 
in  \cite{8845768} by maximization of the SINR term resulting from the UatF bound.

As in \cite{8845768} all RU-UE pairs are associated, the original LSFD scheme for cell-free systems assumes that the elements $\EE[\Gm_k(:,j) \Gm_k(:,j)^\herm]_{m,n}$ are known for $m,n \in [ L ]$ and all $k,j \in [K]$, i.e., the matrix $\EE[\Gm_k(:,j) \Gm_k(:,j)^\herm]$ would have dimension $L \times L$ and have only non-zero entries. 
The scalable LSFD scheme proposed in \cite{demir2021cellfree} in combination with dynamic cooperation clustering assumes that the elements $\EE[\Gm_k(:,j) \Gm_k(:,j)^\herm]_{m,n}$ are known for $m,n \in \Cc_k $ and $k,j \in \Uc(\Cc_k)$. After removing the elements belonging to RU $\ell' \notin \Cc_k$ (equal to zero), the matrix $\EE[\Gm_k(:,j) \Gm_k(:,j)^\herm]$ is of dimension $|\Cc_k| \times |\Cc_k|$ and has only non-zero entries. 

We notice that the LSFD scheme require knowledge of the expected values $\EE[\vv_{m,k}^\herm \hv_{m,j}  (\vv_{n,k}^\herm \hv_{n,j})^\herm]$ for all $m,n \in [L]$, $k,j \in [K]$, and all $m,n \in \Cc_k$, $k,j \in \Uc(\Cc_k)$, respectively. We argue that while the instantaneous values of these coefficients are easily 
obtained from the partial CSI available at the cluster processors, these average values require some sort of long-term averaging 
and thus additional implementation efforts. Therefore, we believe that the scheme proposed in \cite{goettsch2021impact} is not only more performant, but also easier to implement in practice.


For the DL, a popular scheme is local partial ZF (LPZF) proposed in \cite{9069486}. In this scheme, each RU computes the precoding vectors and power allocation locally for its associated UEs. Let us consider the channel matrix between RU $\ell$ and the associated UEs in $\Uc_\ell$, given by
\begin{equation}
	\Hm_\ell = [\hv_{\ell,k_1} \ \hv_{\ell,k_2} \dots \hv_{\ell,k_{|\Uc_\ell|}} ] \in \CC^{M \times | \Uc_\ell | },
\end{equation}
where $k_1, \dots, k_{|\Uc_\ell|}$ are the UEs in the set $\Uc_\ell$. 
In case $M \geq \tau_p \geq |\Uc_\ell|$ and $\Hm_\ell$ is a full rank matrix, local ZF (LZF) is carried out by computing the pseudoinverse
\begin{eqnarray}
	\Hm_\ell^{+} = \Hm_\ell \left ( \Hm_\ell^{\herm} \Hm_\ell \right )^{-1} \label{local_zf} 
\end{eqnarray} 
of $\Hm_\ell$. 
Then, the LZF precoding vector $\uv_{\ell, k}$ is the normalized column of $\Hm_\ell^{+}$ corresponding to user $k \in \Uc_\ell$. 

LPZF is a variant of LZF where some users are excluded from the calculation of the pseudo-inverse. In particular, 
when $M \leq |\Uc_\ell|$, or that $\Hm_\ell$ is rank-deficient ($\tau_p \geq |\Uc_\ell|$ due to clustering), the RU chooses from $\Uc_\ell$ the UEs $\Uc_\ell^{\rm ZF}$ with the largest channel gains (at most $M$) whose channels are linearly independent, and thus form a full-rank matrix $\Hm_\ell^{\rm ZF}$. The precoding vectors of UEs in $\Uc_\ell^{\rm ZF}$ are computed by ZF as in (\ref{local_zf}) and normalized to unit norm. For the remaining UEs $k \in \Uc_\ell^{\rm MRT}$, normalized MRT is employed, i.e., 
\begin{equation}
	\uv_{\ell,k} = \frac{\hv_{\ell,k}}{\| \hv_{\ell,k} \|}, \ \forall k \in \Uc_\ell^{\rm MRT},
\end{equation}
where $\Uc_\ell^{\rm ZF} \bigcap \Uc_\ell^{\rm MRT} = \emptyset$ and $\Uc_\ell^{\rm ZF} \bigcup \Uc_\ell^{\rm MRT} = \Uc_\ell$.

For both cases, the RUs compute the Tx power for each UE locally. 
In conjunction with ZLF and LPZF two simple schemes are considered: equal power allocation (EPA), where the power allocated
to stream $k$ by RU $\ell \in \Cc_k$ is the same for all streams, i.e., 
\begin{eqnarray}
	q_{\ell, k} = \frac{P^{\rm RU}}{| \Uc_\ell |}, \ \forall k \in \Uc_\ell , \label{eq:epa}
\end{eqnarray}
and proportional power allocation (PPA) with regard to the LSFCs such that
\begin{eqnarray}
	q_{\ell, k} = P^{\rm RU} \frac{\beta_{\ell,k}}{ \sum_{j \in \Uc_\ell} \beta_{\ell, j} }, \ \forall k \in \Uc_\ell , \label{eq:ppa}
\end{eqnarray}
where $q_{\ell, k}$ and $P^{\rm RU}$ denote the transmit power allocated at RU $\ell$ to UE $k$ and the DL power budget at each RU, respectively. 
Obviously, in all cases, for $k \notin \Uc_\ell$ we have $q_{\ell, k} = 0$. 

\section{CSI estimation from UL pilots}  \label{CSI-est}

In practice, ideal partial CSI is not available and the channels $\{\hv_{\ell,k} : (\ell,k) \in \Ec\}$ must be estimated from UL pilots. 
Thanks to channel reciprocity in the TDD mode, the estimates can be used for both UL combining and DL precoding.
We assume that $\tau_p$ signal dimensions per RB are dedicated to UL pilots (see \cite{3gpp38211}), and define a codebook of $\tau_p$ orthogonal pilot 
sequences. 
The pilot field received at RU $\ell$ is given by the $M \times \tau_p$ matrix 
\begin{equation} 
	\Ym_\ell^{\rm pilot} = \sum_{i=1}^K \hv_{\ell,i} \phiv_{t_i}^\herm + \Zm_\ell^{\rm pilot} \label{Y_pilot}
\end{equation}
where $\phiv_{t_i}$ denotes the pilot  vector of dimension $\tau_p$ used by UE $i$ at the current slot. Since the pilot vectors make use of $\tau_p$ symbols, their energy is $\tau_p \SNR$, i.e.,  $\| \phiv_{t_i} \|^2 = \tau_p\SNR$ for all $t_i \in [\tau_p]$. For all UEs $k \in \Uc_\ell$, RU $\ell$ produces the ``pilot matching'' channel estimates
\begin{eqnarray} 
	\widehat{\hv}^{\rm pm}_{\ell,k} & = & \frac{1}{\tau_p \SNR} \Ym^{\rm pilot}_\ell \phiv_{t_k}  \\
	& = & \hv_{\ell,k}  + \sum_{i \neq k : t_i = t_k} \hv_{\ell,i}  + \widetilde{\zv}_{t_k,\ell}   \label{chest}
\end{eqnarray} 
where $\widetilde{\zv}_{t_k,\ell}$ is $M \times 1$ Gaussian i.i.d. with components $\Cc\Nc(0, \frac{1}{\tau_p\SNR})$. Assuming that the subspace information $\Fm_{\ell,k}$ of all $k \in \Uc_\ell$ is known, we consider also the ``subspace projection'' (SP) pilot decontamination scheme
for which the projected channel estimate is given by the orthogonal projection of $\widehat{\hv}^{\rm pm}_{\ell,k}$ onto the subspace spanned by the columns of 
$\Fm_{\ell,k}$, i.e., 
\begin{align}
	\widehat{\hv}^{\rm sp}_{\ell,k} & = \Fm_{\ell,k}\Fm_{\ell,k}^\herm \widehat{\hv}^{\rm pm}_{\ell,k} \\
	& = \hv_{\ell,k}  + \Fm_{\ell,k}\Fm_{\ell,k}^\herm \left ( \sum_{i \neq k : t_i = t_k} \hv_{\ell,i} \right ) + \Fm_{\ell,k} \Fm_{\ell,k}^\herm \widetilde{\zv}_{t_k,\ell}   \label{chest1}
\end{align}
Notice that after the projection, the resulting estimation noise is correlated since it is contained in the channel subspace, 
in fact the covariance matrix of the projected noise is given by $\frac{1}{\tau_p\SNR} \Fm_{\ell,k} \Fm_{\ell,k}^\herm$. 

Writing explicitly the pilot contamination term after the subspace projection, we have 
\begin{align}
	&\Fm_{\ell,k}\Fm_{\ell,k}^\herm \left ( \sum_{i \neq k : t_i = t_k} \hv_{\ell,i} \right ) \nonumber \\ & =  \sum_{i \neq k : t_i = t_k} \Fm_{\ell,k}\Fm_{\ell,k}^\herm \hv_{\ell,i}  \\
	& =  \sum_{i \neq k : t_i = t_k} \sqrt{\frac{\beta_{\ell,i} M}{|\Sc_{\ell,i}|}}  \Fm_{\ell,k}\Fm_{\ell,k}^\herm \Fm_{\ell,i} \nuv_{\ell,i} .
\end{align}
This is a $M \times 1$ Gaussian vector with mean zero and covariance matrix
\[ \Sigmam_{\ell,k}^{\rm co}  = \sum_{i \neq k : t_i = t_k} \frac{\beta_{\ell,i} M}{|\Sc_{\ell,i}|}  \Fm_{\ell,k} \Fm_{\ell,k}^\herm \Fm_{\ell,i} \Fm^\herm_{\ell,i} \Fm_{\ell,k} \Fm^\herm_{\ell,k} .\]
When $\Fm_{\ell,k}$ and $\Fm_{\ell,i}$ are nearly mutually orthogonal, i.e. $\Fm_{\ell,k}^\herm \Fm_{\ell.i} \approx \zerov$,
the subspace projection is able to significantly reduce the pilot contamination effect.

Note that the previously described UL and DL schemes, and the computation for UL-DL duality with ideal partial CSI can be carried out with channel estimates by replacing 
the ideal partial CSI $\{ \hv_{\ell,k} : (\ell,k) \in \Ec\}$ with the channel estimates $\{\widehat{\hv}^{\rm pm}_{\ell,k} : (\ell,k) \in \Ec\}$ or $\{\widehat{\hv}^{\rm sp}_{\ell,k} : (\ell,k) \in \Ec\}$. 
In practical systems, knowledge of the subspace $\Fm_{\ell,k}$ for $(\ell,k) \in \Ec$ however is not directly available and requires non-trivial estimation. 

How to efficiently estimate the user channel covariance or at least their dominant signal subspace without suffering from the same pilot contamination 
that appears in the channel estimate themselves is a very relevant and interesting open problem.
Notice also that most papers and monographs on the subject (e.g., see \cite{9336188} and references therein) assume that all channel statistics (in particular, the channel covariance matrices) are ``magically'' known everywhere. This assumption is extremely unrealistic since although the channel statistics change quite slowly in time with respect to the small-scale fading coherence time, they are generally time-varying and must be continuously tracked in order to enable 
schemes such as the subspace projection or some forms of MMSE pilot decontamination as proposed in  \cite{9336188} to work properly.

\section{Numerical results}

In our simulations, we consider a square coverage area of $A = 225 \times 225$ square meters with a torus topology to avoid boundary effects. 
The LSFCs are given  according to the 3GPP urban microcell pathloss model from \cite{3gpp38901}. 
With $N_0=-96 \text{dBm}$, the UL power $P^{\rm ue}$ is chosen such that $\bar{\beta} M \SNR = 1$ (i.e., 0 dB), when the expected pathloss 
$\bar{\beta}$ (averaged with respect to the LOS/NLOS probability and shadowing) is calculated for distance $3 d_L$, where 
$d_L = 2 \sqrt{\frac{A}{\pi L}}$ is the diameter of a disk of area equal to $A/L$. 
We consider RBs of dimension $T = 200$ symbols.  The UL (same for DL) SE for UE $k$ is given by  
\begin{equation}
	{\rm SE}^{\rm ul}_k =  (1 - \tau_p/T) R_k^{\rm ul}.
\end{equation}
The angular support $\Sc_{\ell,k}$ contains the DFT quantized angles (multiples of $2\pi/M$) falling inside an interval of length $\Delta$ placed symmetrically around the direction joining UE $k$ and RU $\ell$. We use $\Delta = \pi/8$ and the maximum cluster size $Q = 10$ (RUs serving one UE) in the simulations. 
The SNR threshold $\eta=1$ makes sure that an RU-UE association can only be established, when $\beta_{\ell,k} \geq \frac{\eta}{M \SNR }$.

For each set of parameters we generated 50 independent layouts (random uniform placement of RUs and UEs), 
and for each  layout we computed the optimistic ergodic rate by Monte Carlo averaging with respect to the channel vectors. 
In all figures, the results with subspace projection channel estimates are shown. 
We compare the DL schemes CLZF and LMMSE with a current state-of-the-art method represented by the LZF scheme of \cite{9069486} (see Section \ref{sec:other-schemes}) as benchmark.

\begin{figure}[t]
	\centerline{\includegraphics[width=.8\linewidth]{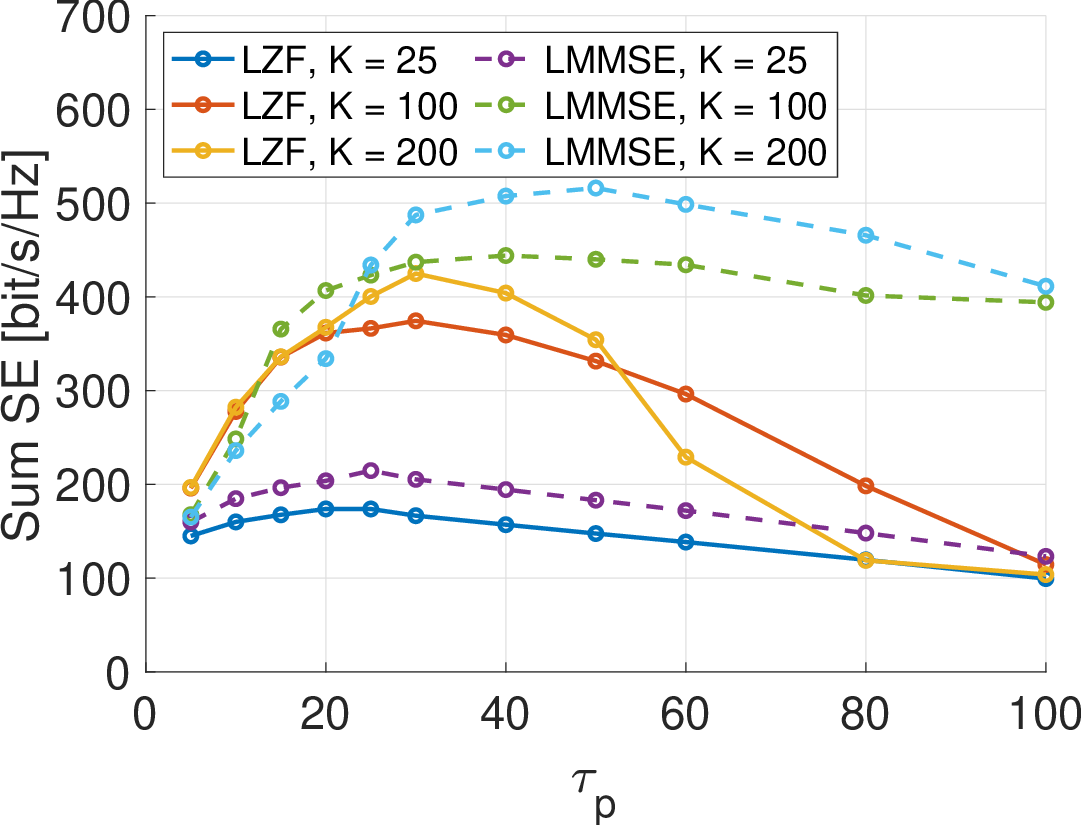}} 
	\caption{Sum DL spectral efficiency vs. $\tau_p$ for different numbers of users $K$ for $L=10$ RUs and $M=64$ antennas each. 
	LMMSE with power allocation from duality, LZF with PPA.}
	\label{fig:compare_K_taup_L10}
\end{figure}

\begin{figure}[t]
	\centerline{\includegraphics[width=.8\linewidth]{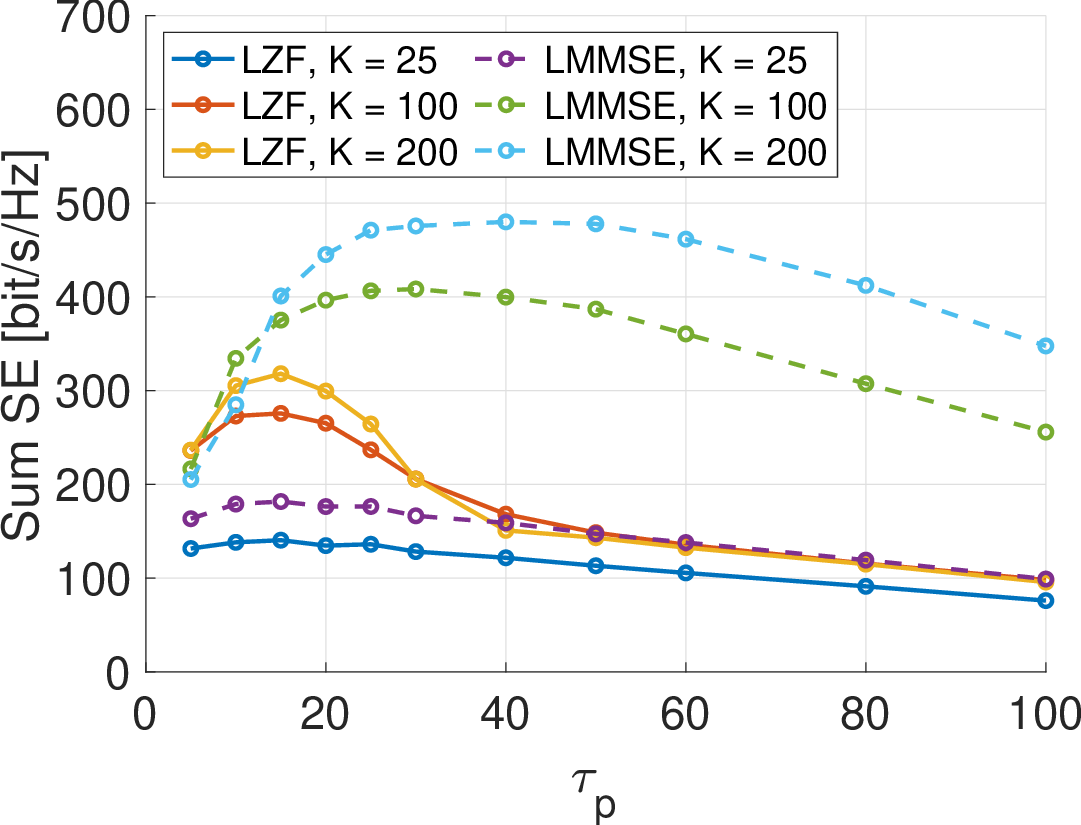}} 
	\caption{Sum DL spectral efficiency vs. $\tau_p$ for different numbers of users $K$ for $L=20$ RUs and $M=32$ antennas each. 
	LMMSE with power allocation from duality, LZF with PPA.}
	\label{fig:compare_K_taup_L20}
\end{figure}

\begin{figure}[t]
	\centerline{\includegraphics[width=.8\linewidth]{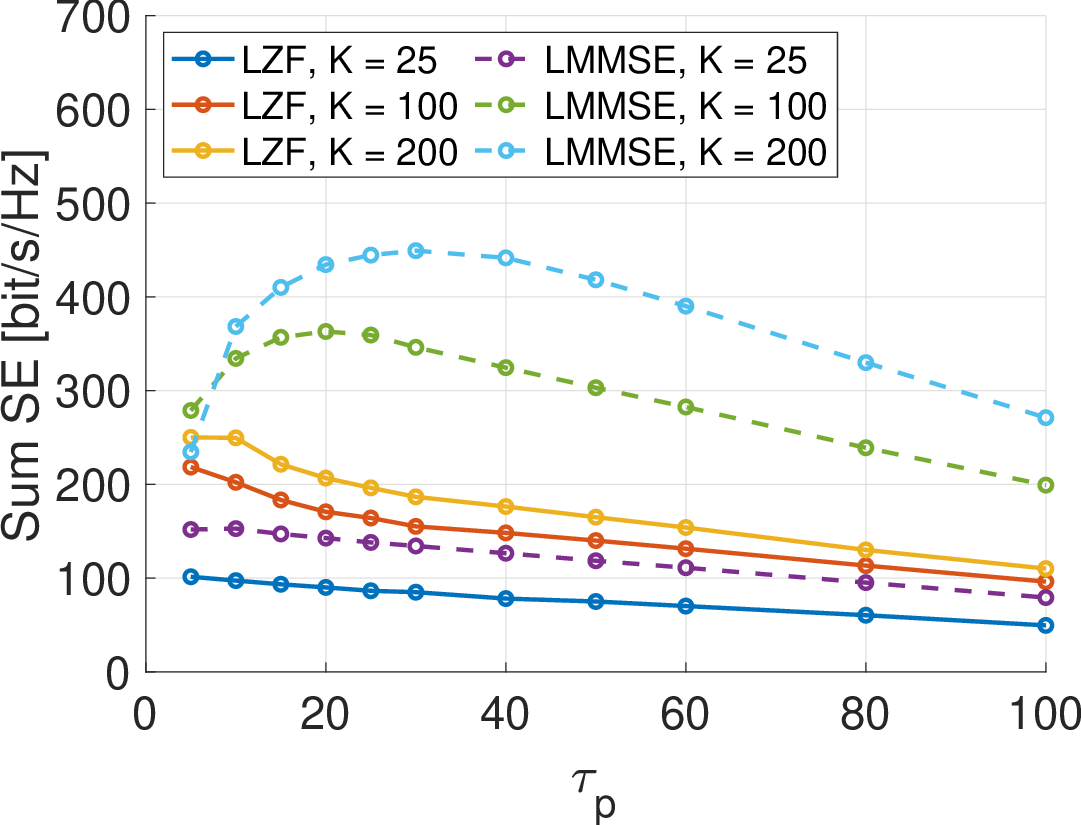}} 
	\caption{Sum DL spectral efficiency vs. $\tau_p$ for different numbers of users $K$ for $L=40$ RUs and $M=16$ antennas each. 
	LMMSE with power allocation from duality, LZF with PPA.}
	\label{fig:compare_K_taup_L40}
\end{figure}

We start our evaluation by comparing the sum SE for different $K$ and $\tau_p$ in a system with a total of $LM = 640$ antennas, 
with different arrangements of $L = 10$, $L = 20$ and $L=40$ RUs (respectively, in Figs.~\ref{fig:compare_K_taup_L10}, \ref{fig:compare_K_taup_L20}, and \ref{fig:compare_K_taup_L40}). 
We notice that in most cases LMMSE outperforms LZF, and that for more UEs in the system, we need larger $\tau_p$ to maximize the sum SE, as each RU can serve up to $\tau_p$ UEs.  We notice also that the SE for the case of $K = 200$ users, at the optimal value of $\tau_p$, is quite invariant with respect to the 
antenna distribution, and in fact the more concentrated antenna distribution ($L = 10$ RUs with $M = 64$ antennas each) yields slightly larger maximum sum SE. 
In any case, this means that the system is quite flexible (within a reasonable range of parameters) with respect to the antenna distribution, and that
a realistic number of RUs with a relatively large number of antennas each represents a practically attractive option with respect to the classical cell-free
paradigm, where the number of RUs is predicated to be larger than the number of users. 

\begin{figure}[t]
	\centerline{\includegraphics[width=.8\linewidth]{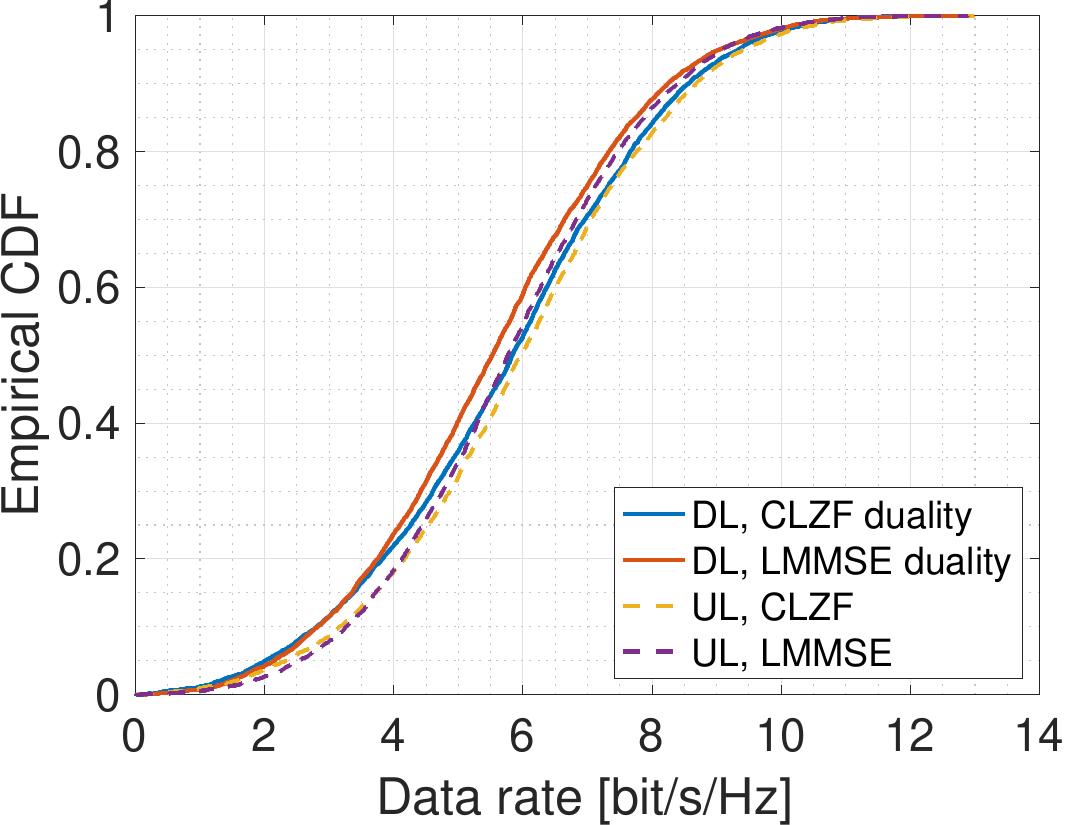}} 
	\caption{Empirical cdf of the UL and DL per-user data rate (in bit/channel use) where $L=10$, $M=64$, $\tau_p=40$.}
	\label{fig:rates_cdf_L10}
\end{figure}

We now look at the distribution of the DL rates per UE for the case $K=100$ users. 
Fig.~\ref{fig:rates_cdf_L10} shows that the proposed UL-DL duality method yields almost symmetric effective ergodic rates for the UL and DL. Also, 
this figure shows that the  CLZF and LMMSE methods perform very similarly. Therefore, since the local LMMSE with cluster-level combining is significantly less 
computationally intensive, it is definitely our preferred and recommended choice. 

\section{Conclusions}

In this paper we have reviewed the basic model for a scalable cell-free user-centric wireless network architecture, 
where each UE is served by a cluster of RUs selected dynamically in the vicinity. In our model, RUs have multiple antennas and are capable of local 
processing. Further processing can be performed at DU nodes at the cluster level. In general, cluster-level decoders are virtualized and are implemented as software-defined network functions, dynamically allocated to DUs in order to maintain a balanced computation load. 
The scalability of the architecture stems from the fact that, provided that the densities of UEs, RUs, and DUs are constant, 
the data rate and computation load at any point of the network remains finite while the network coverage area grows to infinity. 
In a typical practical operational regime, we expect such networks to have DU, RU and UE densities ordered as
$\lambda_{\rm du} < \lambda_{\rm ru} < \lambda_{\rm ue}$, although the number of antennas should be larger than the number of users, 
i.e., $\lambda_{\rm eu} < M \lambda_{\rm ru}$ where $M$ is the number of antennas per RU. 

Following the recent results in  \cite{goettsch2021impact,WCNC2021}, we presented two types of UL linear processing. 
The cluster-level ZF at the cluster for user $k$ computes a receiving vector that sets to zero the interference of all users $j \neq k$ whose channel vectors can be partially estimated by some RU forming the cluster. As an alternative, linear MMSE can be applied separately for each RU, producing a local MMSE estimate for the useful symbol of user $k$. These estimates are then combined by the cluster processor, in order to maximize the SINR of the channel ``after local MMSE estimation'', as seen collectively from all RUs forming the cluster. This second option is much less computationally expensive, and our numerical results show that it provides excellent performance. 

We have also shown that under the partial CSI knowledge available at each given cluster processor there exists a notion of ``nominal'' SINR for which
UL-DL duality holds. This motivates the use of the UL receive vectors also as (dual) DL precoding vectors. We have verified numerically that, despite the
``nominal SINR'' not corresponding exactly to the SINR appearing in the ergodic rate expressions, the rates achieved in UL and DL with this approach are virtually identical. The usage of UL receive vectors as DL precoders has also the non-trivial advantage of dramatically reducing the computation load, since only a set of vectors 
needs to be computed after each new CSI estimation, for both UL and DL. 

Finally, we have proposed a simple subspace projection method to (partially) decontaminate the UL pilots, that exploits the fact that 
with high probability co-pilot users generate channel vectors with different and nearly mutually orthogonal dominant subspaces. 
In fact, by properly assigning the UL pilots to the users, it is unlikely that two users sharing the same pilot (and therefore, by construction, being served by two disjoint clusters of RUs)  are received both at high signal level at a given RU and under the same (or very close) angle of arrival. 
In  \cite{goettsch2021impact} it is shown that the pilot projection method is very effective in closely approaching the system performance under the ideal
partial CSI assumption, that  provides a performance upper bound under the assumption of cluster-level processing.

Several open problems remain to be investigated. In particular, we mention here the design of an effective scheme for estimating the user channel dominant subspace (to implement the pilot projection scheme), and the careful consideration of the mechanisms of UE-RU association and cluster formation. 
In particular, in such type of network architectures, a scheme that allows the UEs to detect the presence of the surrounding RUs and 
a random access procedure to establish the association with at least the cluster leader must be devised. 
In this respect, an interesting option for possible further investigation consists of letting the UEs send ``on-demand'' broadcast signals to be identified by the RUs.
For example, this may happen at the first time a UE joins the system, and may be repeated if the UE detects that its signal quality is below some minimum service threshold, e.g., due to the fact that the user-centric dynamic cluster is not able to evolve rapidly enough to follow the UE mobility. 
A UE-triggered association mechanism with embedded pilot collision detection is presented for example in 
\cite{bursalioglu2018fog}. The scheme in \cite{bursalioglu2018fog} is used at each slot, for a very low-latency UL-DL cycle, but it does not lead to high spectral efficiency, and a similar scheme could be
used only for the initial association and re-association phase. Also, the RUs could apply some form of activity detection and user identification, along the lines 
of massive random access, as for example treated in \cite{fengler2021non}. A complementary option (to be considered as an alternative or together with a suitable random access mechanism) consists of operating the cell-free  user-centric network not in a stand-alone mode, 
but as a a data rate extension - carrier aggregation option to an existing cellular network operating in a different frequency band. 
In this case, the coordination operations such as association and cluster formation could be implemented via a control channel handled by the cellular network, 
These system aspects are left as very promising topics for further investigation.



\fontsize{10}{12}\selectfont
\bibliographystyle{IEEEtran}
\bibliography{massive-MIMO-references}

\end{document}